\documentclass[Letter,useAMS,usenatbib]{mn2e}

\usepackage{graphicx}
\usepackage{hyperref}
\usepackage{lscape}
\usepackage{amsmath}
\usepackage{amsfonts}
\usepackage{amssymb}
\usepackage[usenames,dvipsnames]{color}
\usepackage{float}
\usepackage{ragged2e}
\usepackage[section]{placeins}

\begin{document}

\title[]{The impulsive phase of magnetar giant flares: assessing linear tearing as the trigger mechanism}

\author[Elenbaas et al.]{C. Elenbaas$^1$\thanks{E-mail: C.P.C.Elenbaas@uva.nl},  A. L. Watts$^1$, R. Turolla$^{2,3}$, J. S. Heyl$^4$ 
\\
  $^{1}$Anton Pannekoek Institute for Astronomy, University of
  Amsterdam, Science Park 904, 1098XH, Amsterdam, The Netherlands.\\
  $^{2}$Department of Physics and Astronomy, University of Padova, via Marzolo 8, 35131, Padova, Italy.\\
  $^{3}$Mullard Space Science Laboratory, University College London, Holbury St. Mary,Surrey, RH5 6NT, UK .\\
  $^{4}$University of British Columbia, Vancouver, Canada.}

\pagerange{\pageref{firstpage}--\pageref{lastpage}}
\pubyear{2014}

\maketitle

\label{firstpage}

\begin{abstract}
Giant $\gamma$-ray flares comprise the most extreme radiation events observed from magnetars. Developing on (sub)millisecond timescales and generating vast amounts of energy within a fraction of a second, the initial phase of these extraordinary bursts present a significant challenge for candidate trigger mechanisms. Here we assess and critically analyse the linear growth of the relativistic tearing instability in a globally twisted magnetosphere as the trigger mechanism for giant $\gamma$-ray flares. Our main constraints are given by the observed emission timescales, the energy output of the giant flare spike, and inferred dipolar magnetic field strengths. We find that the minimum growth time of the linear mode is comparable to the $e$-folding rise time, i.e. $\sim10^{-1}$ ms. With this result we constrain basic geometric parameters of the current sheet. We also discuss the validity of the presumption that the $e$-folding emission timescale may be equated with the growth time of an MHD instability.
\end{abstract}


\section{Introduction}\label{sec:intro}

Magnetars are neutron stars (NSs) whose output power is dominated by the decay of an ultra-strong magnetic field (often exceeding the quantum critical field, $B_{\rm qed}\equiv m_e^2c^3/(e\hbar)\simeq4.41\times10^{13}$ G) \citep[\citealt{Thompson95}, e.g.][]{Mereghetti08,Turolla15}. The transient emission properties of such sources include comparatively minor recurrent soft $\gamma$-ray bursts ($E\lesssim10^{42}$ erg) and sporadic giant $\gamma$-ray flares ($E \sim 10^{44}-10^{46}$ erg)\footnote{Energy discharge estimates assume an isotropic release of radiation.}. At present, three giant flares have been observed from independent sources and their lightcurves exhibit remarkably similar characteristics (see Figure~\ref{fig:lcurve} in Section \ref{sec:et}). Giant flares are typically composed of an explosive initial hard $\gamma$-ray spike ($k_{\rm B}T_{\rm spec}\sim175-250$ keV) that develops within (sub)milliseconds and lasts a mere fraction of a second ($\sim0.15-0.5$ s), and a quasi-exponentially abating x-ray tail ($\sim20-30$ keV) that persists for minutes, with superimposed pulsations \citep[see e.g.][]{Mazets79,Fenimore96,Hurley99,Feroci01,Palmer05} . 

The emission of the decaying tail is argued to be the result of a continuously evaporating and locally magnetically trapped thermal photon-pair fireball. Beamed emission from this moves in and out the line of sight, due to the rotation of the underlying NS \citep{Thompson95}. The physical process behind the onset, the trigger mechanism, that would clarify the impulsive phase of these energetic flares, remains however a topic of great debate. Here we will discuss one such mechanism, spontaneous tearing of a globally extended equatorial current sheet, in more detail. Typical emission timescales of the observed giant flares play a critical role in resolving this dispute.

\subsection{Giant flare trigger mechanisms}

In this section we briefly explore the various magnetar giant flare trigger mechanisms that have been proposed. We begin with the setup of the system prior to the explosive event and proceed with the triggers, subdivided in internal and external mechanisms.

\subsubsection{Setup: magnetic field formation and evolution}

The origin of the strong magnetic field is a non-trivial affair. \citet{Thompson92} have argued that during the transient phase of extensive neutrino cooling moments after gravitational collapse of the progenitor star, entropy-driven convection and differential rotation inside a rapidly spinning (initial spin period: $\Omega_0^{-1}\sim1$ ms) proto-NS may sustain an efficient $\alpha$-$\Omega$ dynamo which could generate an internal magnetic field up to $\sim10^{17}$ G. Alternatively, the massive progenitor may already accommodate a sizeable magnetic field. An ultra-strong field is consequently formed via straightforward flux conservation of the fossil field during implosion \citep{Ferrario06}.

The dynamical timescale of the newly formed ultra-strong field is only seconds or less, and the crystallisation of the outer layer does not set in for another couple of minutes to hours. This allows the field to evolve readily towards a (meta-)stable magneto-hydrodynamic (MHD) equilibrium configuration, likely consisting of a combination of a poloidal- and toroidal component, before its further evolution is constrained by the presence of a highly conductive solid crust \citep{Flowers77,Braithwaite06}. The problem of magnetic-field stability, and the respective strengths of the two field components have been studied by e.g. \citet{Braithwaite09}, \citet{Lander12}, and \citet{Ciolfi13}. No consensus has been reached on these matters yet, and further investigations including the effects of superconductivity \citep{Lander14, Henriksson13} and the NS crust \citep{Gourgouliatos14} are required to advance the issue. Subsequent evolution of the strongly twisted field is then determined by ambipolar diffusion, Ohmic decay, and (non-diffusive) Hall drift which occur throughout the interior of the NS (crust and core) and operate on much longer timescales $\gtrsim10^4$ yr \citep{Thompson96,Heyl98}. The conductive crust either severely resists the imparted motion of the frozen-in magnetic flux tubes such that Maxwell stress builds up continuously in the system or allows for a constrained transport of magnetic helicity into the magnetosphere, which in turn may develop into a sheared configuration. A reservoir of energy grows (internally or externally) until a certain critical threshold is reached, suddenly releasing the energy in an explosive manner through e.g. a crustal failure or MHD instability of the magnetic field.

\subsubsection{Internal trigger}

\begin{figure*}
	\centering
		\includegraphics[width= 0.9 \textwidth]{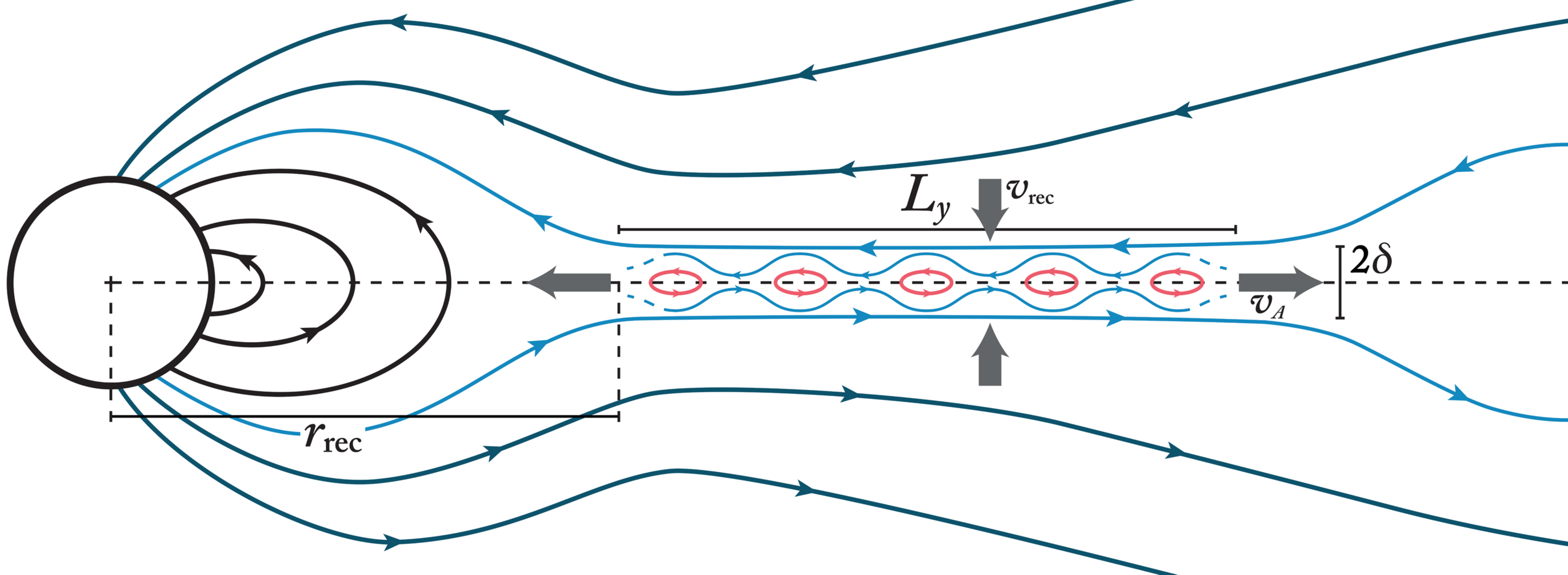}
	\caption{2D cut-through of the globally sheared magnetic field containing an equatorial current sheet. Continuous current injections into the magnetosphere gradually increase the magnetic helicity of the external field. This in turn may evolve continuously towards a Y-type neutral line configuration, such that a narrow current sheet is formed. Reconnection through spontaneous tearing of the current sheet results in magnetic field dissipation and the ejection of a relativistic magnetic plasmoid. The dimensions of the current sheet have been labeled (current sheet length $L_y$ and thickness $2\delta$) and the height of the base of the reconnection region with respect to the centre of the NS is given by $r_{\rm rec}$. The velocities $v_{\rm rec}$ and $v_A$ are the inward reconnection and outward (Alfv\'enic) bulk plasma velocity, respectively.}\label{fig: mag_sphere1}
\end{figure*}

Motivated by the duration of the impulsive phase ($\sim0.1-1$ s) \citet{Thompson95} initially proposed an internal trigger mechanism whereby a large-scale interchange instability, i.e. a global MHD rearrangement, would take place in the liquid core of the NS and propagate outward on a dynamical timescale, equal to the internal Alfv\'en crossing time, 
\begin{equation}\label{eq:int. Alfven cross time}
\tau^{\rm int}_A=\frac{R_*}{v_A^{\rm int}}\sim0.1~\rm{s},
\end{equation}
where $R_*\sim10^6$ cm is the typical radius of a NS and $v_A^{\rm int}\sim10^7\,B^{\rm int}_{15}$ cm s$^{-1}$ is the core Alfv\'en speed for a density $\sim10^{15}$ g cm$^{-3}$ with the core magnetic field strength given by $B^{\rm int}\equiv B^{\rm int}_{15}\times10^{15}$ G. This results in a sudden global displacement of the magnetic footpoints on the surface of the star injecting an `Alfv\'en pulse' into the magnetosphere, which subsequently induces a relativistic outflow of plasma. The (sub)millisecond rise of the giant flare lightcurve is, they argue, the signature of a reconnection front in the magnetosphere leading the relativistic outflow, which in turn develops on the external Alfv\'en crossing time,
\begin{equation}\label{eq:ext. Alfven cross time}
\tau_A^{\rm ext}=\frac{R_*}{v_A^{\rm ext}}\sim3\times10^{-2}~\rm{ms},
\end{equation}
where $v_A^{\rm ext}\sim c$ is the magnetospheric Alfv\'en speed. Therefore, even though we initially observe the emission from the reconnection front, the trigger nevertheless is given by the onset of the internal instability\footnote{See also the discussion in \citet{Link14} on the feasibility of such an internal MHD instability mechanism.}.

A second trigger mechanism introduced by \citet{Thompson01} involves the force balance between the rigidity of the elastic NS crust and vast magnetic shear-stress, imparted through the anchored magnetic field lines\footnote{This mechanism was discussed earlier by \citet{Thompson95} in explaining the physical process behind the less energetic recurrent soft $\gamma$-ray bursts from magnetars.}. Ultimately, the tension of the strongly twisted magnetic field in the crust will become the dominant force and drive the crustal lattice beyond its critical straining threshold $\theta_{\rm crit}$. As the crust yields, the suppressed magnetic energy is allegedly liberated abruptly through a propagating fracture -- analogous to an earthquake -- producing seismic modes, which in turn couple to magnetospheric Alfv\'en modes via the pinned magnetic field lines \citep{Blaes89}. 

\citet{Thompson01} note however that the storage capacity of elastic energy in the crust
\begin{equation}\label{eq:crust energy}
E^{\rm max}_{\rm elastic}\sim 1.7\times10^{43}\left(\frac{\theta_{\rm crit}}{10^{-2}}\right)^{2}~{\rm erg},
\end{equation}
which depends on its critical yield strain, is insufficient to explain the observed output power of a giant flare ($E\gtrsim10^{44}$ erg). Accordingly, they argued that the crust merely functions as a gate that assists in the storage and discharge of the internal magnetic energy, rather than as the main energy reservoir. It is important to remark however that they assumed conservatively $\theta_{\rm crit}\lesssim10^{-2}$, yet this value has since been revised by \citet{Horowitz09} through molecular dynamics simulations to be $\theta_{\rm crit}\sim0.1$ \citep[this value has been independently reproduced by][]{Hoffman12}. With this we obtain $E^{\rm max}_{\rm elastic}\sim10^{45}$ erg [see Eq.~(\ref{eq:crust energy})], which is comparable to the total energy output of the giant flares. Note however that the value for the critical breaking strain decreases significantly, due to defects induced in the crust after the first time it yields \citep{Hoffman12}. Nonetheless, \citet{Lander15} argue that even a moderate breaking strain of $\sim0.065$ and a fracture extending to the base of the crust can power the most energetic giant flare to date.

Due to the large hydrostatic pressure in the NS crust $P_{\rm crust}$ in comparison to the shear modulus $\mu$, i.e. $P_{\rm crust}\gg \mu$, it is impossible to create a long-lived void necessary for a brittle fracture to occur \citep{Jones03}, regardless of the magnitude of the imparted Maxwell stress. When the crust yields it does not crack, yet rather undergoes a gradual plastic deformation in response to the imparted Lorentz forces, whereby internal currents and associated magnetic helicity are transported outward into the less conductive magnetosphere \citep{Thompson02}. 

\citet{Levin12} argue that the presence of a strong magnetic field reinforces the crust, which might strongly impede the formation of a propagating fracture or global slip, altogether. Only under certain specific conditions, where the magnetic flux surface is oriented almost perfectly perpendicular to the direction of shear (within $10^{-3}$ radian), can enough energy be released through a propagating fracture to explain the observed emission. 

An important challenge for trigger mechanisms that manifest internally, either a core MHD instability or crustal failure, is the significant impedance mismatch between the internal and external Alfv\'en velocities \citep{Link14}. As a result, magnetic energy that dissipates through shear waves cannot be transmitted to the magnetosphere fast enough to explain the (sub)millisecond rise of the initial transient phase of the giant flare. Instead, shear waves get reflected numerous times prior to leaving the stellar interior, extending the outward transmission time considerably.

\subsubsection{External trigger}

The aforementioned issues with internal triggers have led to the notion that prior to a giant flare the magnetic energy might be stored in the magnetosphere, rather than in the interior of the NS. \citet{Thompson02}  argue that the tightly wound internal magnetic field induces a strong current that in turn closes through a thin surface layer. This local surface layer will experience a Lorentz force, which causes the crust to rotate plastically. Anchored magnetic field lines are dragged along with the gyrating motion and a twist is gradually imparted to the external magnetic field. The twist supporting currents can be composed by charges stripped from the NS surface or -- more likely -- by pair creation in the magnetosphere \citep{Beloborodov07}. Subsequently, the non-potential external field reacts to the new boundary conditions and evolves through a series of quasi-equilibria, continuously twisting the external field either locally \citep{Huang14a,Huang14b, Beloborodov09} or globally \citep{Thompson02}.

A local increase of helicity leads to the formation of a helically twisted flux rope embedded in the magnetar magnetosphere, whereby the impulsive phase of the giant flare is associated with an abrupt loss of equilibrium and subsequent catastrophic destabilisation of the flux rope, analogous to  the dynamics of coronal mass ejections (CMEs) \citep{Masada10,Yu12,Yu13,Huang14a,Huang14b}. Alternatively, a global accumulation of twist may cause the external field to eventually expand outwards, becoming increasingly radial, and admitting a cusp-shaped or Y-type neutral line topology, characterised by a narrow equatorial current sheet where the magnetic shear is most significant \citep{Mikic94,Wolfson95,Parfrey13}. In this narrow yet extended neutral layer the gradients become significant and the MHD approximation breaks down allowing for the field lines to diffuse through the plasma. The onset of the flare is then given by an explosive reconnection event, which may roughly develop on the external Alfv\'en crossing time $\tau_A^{\rm ext}\sim10^{-2}$ ms [Eq.~(\ref{eq:ext. Alfven cross time})] \citep{Thompson95}, and the expulsion of a relativistic plasmoid. In this paper we investigate specifically the reconnection process in the latter configuration -- illustrated in Figure \ref{fig: mag_sphere1}.

Both magnetospheric models provide a mechanism for slow build up of an energy reservoir over tens of years caused by the ambipolar diffusion of the internal magnetic field and its subsequent rapid conversion into bulk kinetic energy, particle acceleration, and radiation ($\lesssim$ milliseconds). Observed spectral hardening (softening) and an increase (decrease) in spin down in the pre (post) giant flare stage of SGR 1900+14 and SGR 1806-22 \citep{Woods99,Woods01,Mereghetti05, Rea05} are consistent with an increase (decrease) of twist and charge density in the external field \citep{Thompson02,Lyutikov06}. Moreover, a considerable reduction in harmonic content of the pulse profile of SGR 1900+14 during and following the giant flare suggests a burst mechanism which reduced the twist of the external field significantly \citep{Woods01}.

Distinct reconnection models have been introduced to describe the initial transient phase of the observed giant flares. \citet{Lyutikov03,Lyutikov06} and \citet{Komissarov07} suggest the development of the tearing instability in a relativistic force-free current sheet as the trigger mechanism. They argue that the minimum growth time of the linear tearing mode accords with the (sub)millisecond rise (to peak) of the giant flares. Alternatively, \citet{Gill10} propose a fast reconnection model that relies on collisionless Hall reconnection ($\tau_{\rm rec}^{\rm Hall}\sim0.3$ ms) and ascribes a crucial role to the soft precursors ($\gtrsim10^{41}$ erg, $k_{\rm B}T<50$ keV) that have been observed before the last two giant flares\footnote{
Any precursor of the 1979 March 5 giant flare would have gone by unnoticed due to the lack of detectors operational at the time with sensitivities below $\sim50$ keV.}. These precursors facilitate the conditions for collisionless Hall reconnection by introducing a baryon contaminant in the pair dominated magnetosphere, since the former process relies on the Hall effect which in turn requires a non-mass-symmetric plasma composition to operate.\\
\\
In this article we focus on magnetospheric giant flare trigger mechanisms. In particular we critically analyse the most discussed candidate reconnection mechanism, i.e. impulsive reconnection through the spontaneous development of the tearing instability in a globally sheared external field.\footnote{ Recent particle-in-cell simulations that describe relativistic reconnection in pair plasmas demonstrate the growth of the drift-kink (DK) instability perpendicular to the plane of reconnection through tearing \citep{Zenitani07}. For certain initial equilibrium configurations the DK instability dominates over the tearing instability at first and consequently impedes efficient reconnection, thermalises the particles, and broadens the current sheet. However, it is also shown that efficient reconnection leading to significant particle acceleration will occur at a later stage, when the tearing mode regains dominance \citep{Sironi14}. Moreover, it is found that the DK instability is quenched in the presence of a finite guide field, such that the dynamics of the sheet is dictated by the development of the tearing instability at all stages \citep{Zenitani08,Kagan13,Cerutti14}. The $B_\phi$ component of the globally twisted magnetic field surrounding the neutron star may act as a guide field in the case of an equatorial current sheet.} We revise the tearing mode growth time as applied to magnetar magnetospheres by \citet{Lyutikov03} and expand on the rectified result. Characteristic timescales appearing in the giant flare lightcurves have hereby provided necessary constraints. Furthermore, we provide order of magnitude estimates related to the geometry of the reconnection region and discuss the validity of basic assumptions regarding this trigger mechanism.

In Section \ref{sec:et} we review typical timescales of the observed giant flare emission and additional relevant data of the phenomena involved. In Section \ref{sec:tmt} we summarise previous works on the dynamics behind the relativistic tearing instability and the general expression for its minimum growth time. Subsequently we show that a revised version of the tearing mode growth time for magnetar magnetospheres can in principle explain the (sub)millisecond rise times of giant flares under certain conditions, pertaining to the geometry of the reconnection region. In Section \ref{sec:pcotrr}, using straightforward theoretical models that rely on the tearing mode timescale, we constrain the height of the reconnection region and thickness of the current sheet in which the tearing mode develops. In section \ref{sec:ep} we discuss the relations between the MHD growth time and the radiative timescale, which is directly connected with the observed lightcurve. Throughout the article we adopt a Gaussian-cgs unit system in our calculations. 

\section{Emission Timescales}\label{sec:et}

Currently, three magnetars have produced a giant $\gamma$-ray flare. In chronological order they are the 1979 March 5 flare from SGR 0526-66 \citep{Mazets79}, the 1998 August 27 flare from SGR 1900+14 \citep{Hurley99}, and the 2004 December 27 flare from SGR 1806-20 \citep{Palmer05}. The energy in radiation emitted during the decaying tail was roughly equal for the three giant flares ($E_{\rm tail}\sim10^{44}$ erg), indicating that the strength of the confining magnetospheric field, which traps the photon-pair fireball, is roughly similar for the three sources since the energy storage capacity of the field is related to its strength \citep{Mereghetti08}. The inferred surface dipole magnetic field strengths $B_s$ of the sources are given in Table \ref{tab:gf data}. The released photon energy during the initial spike was however considerably larger for the most recent giant flare ($E_{\rm spike}\sim10^{46}$ erg), as compared to the first two ($E_{\rm spike}\sim10^{44}$ erg).

Considering the fact that the duration of the hard spike is roughly three orders of magnitude less than the soft tail, it is rather astonishing that the photon energy output of the hard spike is approximately equal to or even much greater than the energy released during the decay of the soft tail. The conversion of such a vast amount of stored magnetic energy into high energy radiation in a considerably limited window of time, requires an extraordinary trigger mechanism indeed, which accordingly may be constrained by the observed photon flux and associated sub(milli)second rise time.

\begin{table*}
\centering
\caption{\small{Observed emission timescales from magnetar giant flares. Including auxiliary parameters: source distance $d$, isotropic peak luminosity $L_{\rm peak}$, spike energy $E_{\rm spike}$, spectral temperature $k_{\rm B}T_{\rm spec}$, and inferred surface dipole magnetic field strength $B_{\rm s}$.}}
\begin{tabular}{llllllll}	
\hline
SGR & & \textbf{0526-66} & & \textbf{1900+14} & & \textbf{1806-20} & \\
\hline
\hline
Date &  &1979 March 5 & & 1998 August 27 & & 2004 December 27 &\\
&&& \emph{Ref.} && \emph{Ref.} && \emph{Ref.}
\\
$\tau_e$ & [ms] & $\lesssim1$ & [4] & $<1.6,<4$ & [14], [6] & $\lesssim0.3,\,<1$ & [13], [7]\\
$\tau_{\rm peak}$ & [ms] & $\sim15$, $\sim20$ & [8], [2] & - & & $\sim1.5$ & [13]\\
$\tau_{\rm spike}$ &[s] & $\sim0.1-0.2$ & [9] & $\sim0.35, \sim1.0$ & [10], [6] & $\sim0.2$, $\sim0.5$ & [7], [13] \\
\\
$d$ & [kpc] & 53.6 & [5]$^{\dagger}$ & 12.5 & [3]$^{\dagger}$ & 8.7 & [1]$^{\dagger}$\\
\\
$L_{\rm peak}$* &[$10^{44}$ erg s$^{-1}$] & $\sim4.7,\,\sim18$** & [8], [4] & $>0.64,\,>5.6,\,>160$ & [6], [10], [14] & $\sim7\times10^2$ & [7]\\
$E_{\rm spike}$* &[$10^{44}$ erg] & $\sim1.1$ & [8] & $>0.10,\,>3.0$ & [10], [14]  & $\sim1.2\times10^2$ & [7]\\
$k_{\rm B}T_{\rm spec}$ & [keV] & $246$ & [4] & 240 & [6] & 175 & [7]\\
\\
$B_{\rm s}$ &[$10^{14}$ G] & 5.6 & [15]$^{\dagger}$ & 7.0 & [11]$^{\dagger}$ & 20 & [12]$^{\dagger}$
\\\hline
\end{tabular}\begin{justify}{\small *Reference peak luminosities and spike energies have been adjusted according to respective source distances in [5], [3], and [1]. **Peak luminosity from [8] ([4]) is an average over 200 (10) ms. \emph{References}: [1] \citet{Bibby08}; [2] \citet{Cline80}; [3] \citet{Davies09}; [4] \citet{Fenimore96}; [5] \citet{Haschke12}; [6] \citet{Hurley99}; [7] \citet{Hurley05}; [8] \citet{Mazets79}; [9] \citet{Mazets81}; [10] \citet{Mazets99}; [11] \citet{Mereghetti06}; [12] \citet{Nakagawa09}; [13] \citet{Palmer05}; [14] \citet{Tanaka07}; [15] \citet{Tiengo09}. $^{\dagger}$References obtained through the `McGill Online Magnetar Catalog' \citep{Olausen14}: \url{http://www.physics.mcgill.ca/~pulsar/magnetar/main.html}. Burst references can also be found at the Amsterdam Magnetar Burst Catalogue: \url{http://staff.fnwi.uva.nl/a.l.watts/magnetar/mb.html}.}\end{justify}\label{tab:gf data}
\end{table*}

\subsection{Timescale definitions}\label{sec:emission timescales}

\begin{figure}
	\centering
		\includegraphics[width=0.46 \textwidth]{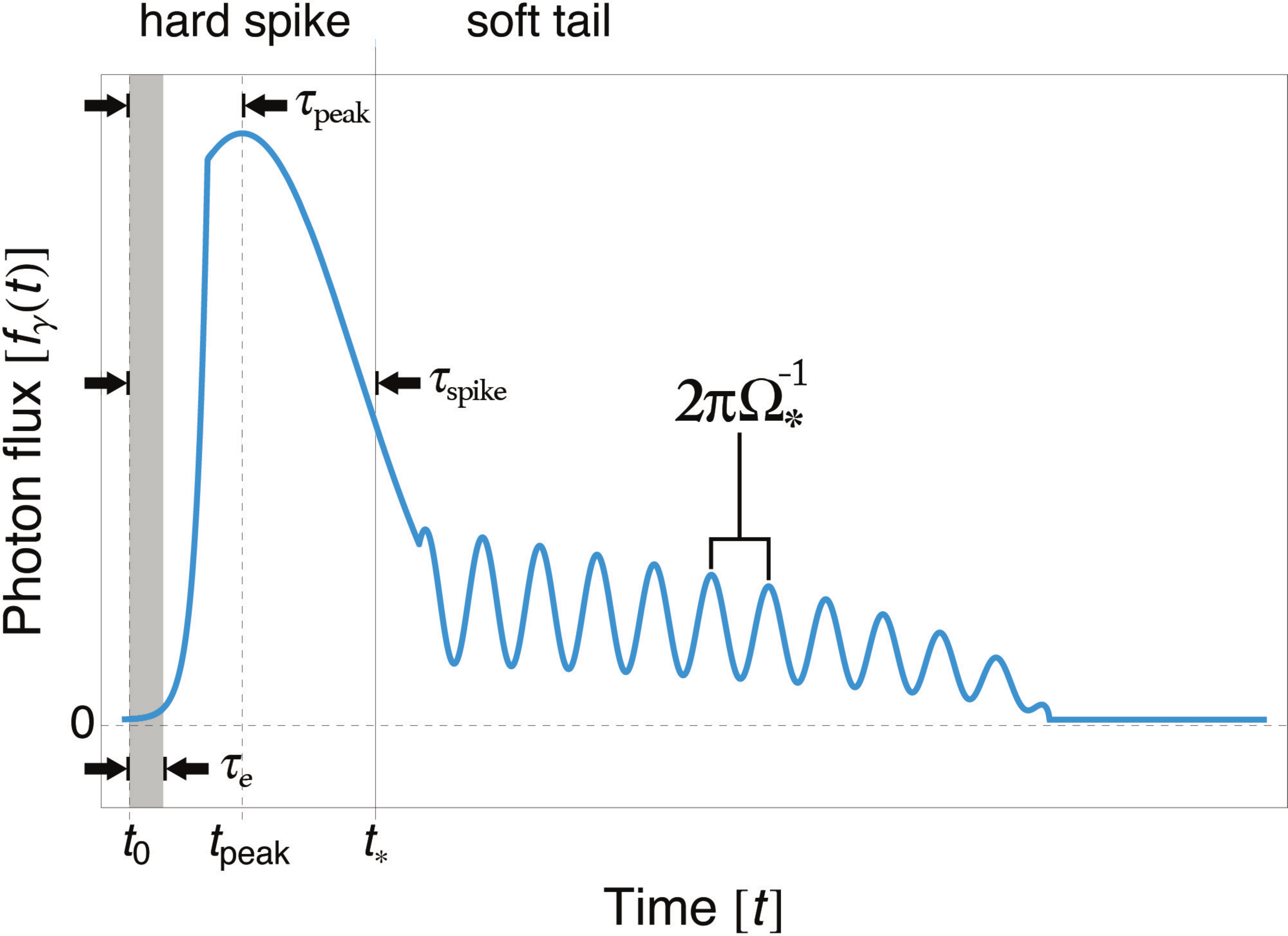}
	\caption{Schematic representation of giant $\gamma$-ray flare lightcurve with the $\gamma$-ray photon flux as a function of time. Typically the giant flare light curve may be subdivided into two regions characterised by their respective spectral hardness: the spectrally hard impulsive phase, i.e. the hard spike, and the spectrally soft afterglow or tail with superimposed pulsations. The onset of the giant flare is at $t_0$, the hard spike reaches its peak flux at $t_{\rm peak}$, and $t_*$ denotes the end of the spectrally hard phase. The grey area denotes the exponential rise timescale of the spike, $f_\gamma\propto\exp(t/\tau_e)$. The peak time is defined as $\tau_{\rm peak}\equiv |t_{\rm peak}-t_0|$ and the spike duration time as $\tau_{\rm spike}\equiv|t_*-t_0|$. Note that, since in reality $\tau_{\rm spike} \ll \tau_{\rm tail}$, the time domain of the initial phase has been magnified for viewing purposes. The most distinct mode of the modulated tail emission has a period equal to the rotation period $2\pi\Omega_*^{-1}$ of the NS.}\label{fig:lcurve}
\end{figure}

In studying the initial spectrally hard phase of the giant flare lightcurve, the following characteristic emission timescales may be defined\footnote{Here we follow the definitions for the typical timescales as described by \citet{Duncan04}, section 1.3.} (see Figure~\ref{fig:lcurve}). The $e$-folding rise time $\tau_e$ describes the exponential rise of the spike out from the continuum ([$f_\gamma\propto\exp(t/\tau_e)]$, where $f_\gamma$ represents the photon flux). This emission timescale constrains the explosive capability of the trigger mechanism, i.e. the physical process that generates the observed radiation is necessarily required to advance on this timescale. The peak time $\tau_{\rm peak}\equiv |t_{\rm peak}-t_0|$ denotes the time between the onset of the spike $t_0$ and the moment $t_{\rm peak}$ when the spike photon flux peaks $[f_\gamma^{\rm max}(t_{\rm peak})]$ and the spike time $\tau_{\rm spike}\equiv|t_*-t_0|$ represents the duration of the spike, i.e. the timespan of the spectrally hard phase of the giant flare lightcurve, where $t_*$ indicates the end time of the hard spike. The latter timescale may serve to constrain the energy deposition or radiative evaporation time. This timescale will depend on factors such as the extent of the energy reservoir, the rate of energy conversion and radiation production, and/or the effective trapping of the generated radiation.

\subsection{Observed characteristic timescales and auxiliary parameters}\label{ssec:observed timescales}

From the giant flare initial spike data listed in Table \ref{tab:gf data}, we find that the values for the various timescales are typically, $\tau_e\sim0.1-1$ ms, $\tau_{\rm peak}\sim1-10$ ms, and $\tau_{\rm spike}\sim0.1-1$ s. However, the accuracy and precision of the $e$-folding rise time measurements is restricted by the limited time-resolution of the detectors operational at the time. Moreover, the short timescales may have been significantly affected by saturation of the detector and deadtime of the instrument. Both effects, if present, result in an overestimation of the shortest timescales and in particular the $e$-folding rise times. Therefore strictly one should regard these timescales as upper limits.

The listed spectral temperatures $k_{\rm B}T_{\rm spec}$ in Table \ref{tab:gf data} are obtained through fitting optically thin thermal Bremsstrahlung (OTTB) or cooling blackbody models to the spectra of the observed giant flare spikes \citep{Fenimore96,Hurley99,Hurley05}. However, the exact physical mechanism that generates the observed spectra remains unknown.

The initial spikes display strong variability on (sub)millisecond timescales and quasi-periodic oscillations (QPOs) with $\nu\sim10^2$ Hz  \citep{Barat83,Hurley99,Feroci01,Terasawa05,Schwartz05}. Peak luminosities and spike energies in Table \ref{tab:gf data} are found assuming isotropic radiation and computed from the observed fluxes using the respective source distances; no spectral bolometric corrections have been applied. In Table \ref{tab:gf data}, multiple values are quoted at times for various quantities. These values have been sourced from distinct references. They differ because of significant differences in instrumentation, e.g. energy bandwidth and time resolution, and in data analysis techniques. We quote these values to give an indication of the uncertainties involved.

\section{The relativistic tearing mode}\label{sec:tmt}

Here we consider the development of the tearing instability in a relativistic current sheet as depicted in Figure~\ref{fig: mag_sphere1}. In the presence of finite magnetic resistivity $\eta$ the current sheet will become unstable to transverse tearing modes ($\mathbf{k}\cdot\mathbf{B}=0$) and decompose into many smaller current filaments or magnetic islands \citep{Furth63} -- see Figure \ref{fig: mag_sphere1}. Simultaneously magnetic energy is converted into heat, bulk kinetic energy, and charged particles are accelerated by the induced reconnection electric field $\mathbf{E}=-E_z\hat{\mathbf{z}}$ \citep[see e.g.][]{Priest00}.

In the following we revisit and further analyse the (relativistic) tearing instability as a candidate trigger mechanism for the onset of magnetar giant flares, the groundwork for which has been laid in detail by \citet{Lyutikov03} and \citet{Komissarov07}. Here we briefly review the relevant equations of resistive magnetodynamics and the stability analysis of a current sheet in a magnetically dominated magnetosphere, which ultimately results in a minimum growth time of the linear tearing mode. Next we discuss the application of this characteristic timescale to the initial rise of magnetar giant flares and reassess the conclusions of previous work. 

\subsection{Force-free degenerate electrodynamics}

\subsubsection{Magnetization parameter}

To investigate the properties of the magnetar magnetosphere it proves useful to define the dimensionless magnetization parameter,
\begin{equation}\label{eq}
\sigma_{\rm m}\equiv2\frac{u_{\rm B}}{u_{\rm p}}=\frac{B^2}{4\pi\rho c^2},
\end{equation}
which describes the ratio of magnetic energy density to total particle energy density, where $u_{\rm B}=B^2/8\pi$ and $u_{\rm p}=\rho c^2$, with $B$ the magnitude of the magnetic field, $\rho$ the particle density, and $c$ the speed of light. The magnetization parameter for magnetar magnetospheres is estimated to be  $10^{13}\leq\sigma_{\rm m}\leq10^{16}$ \citep{Komissarov07}. When $\sigma_{\rm m}\gg1$, the magnetosphere is said to be magnetically dominated (the inertia of the particles is negligible, even though they still act as carriers of charge) and relativistic, since the velocity of an Alfv\'en wave,
\begin{equation}\label{eq}
v_A= c\left(\frac{\sigma_{\rm m}}{1+\sigma_{\rm m}}\right)^{1/2},
\end{equation}
approaches the speed of light, i.e. $v_A\to c$. Note accordingly that the Alfv\'en transit time becomes the light crossing time, $\tau_A\to\tau_c=l/c$, where $l$ denotes the typical length scale of the system. 

In describing the dynamics of the magnetar magnetosphere, $\sigma_{\rm m}^{-1}$ may be used as a small expansion parameter to approximate the general equations of relativistic magnetohydrodynamics (RMHD) in the limit of vanishing rest-mass density and pressure of matter (force-free approximation), i.e. force-free degenerate electrodynamics (FFDE) or `magnetodynamics' (MD) \citep{Uchida97,Komissarov02,Komissarov07}.

\subsubsection{Ohm's law in resistive FFDE}

In FFDE the energy-momentum equation in covariant form, stripped from its matter component, reduces to
\begin{equation}\label{eq:energy-mom}
\nabla_{\mu}T_{\rm em}^{\mu\nu}=0,
\end{equation}
where
\begin{equation}\label{eq}
T_{\rm em}^{\mu\nu}=\frac{1}{4\pi}\left[F^{\nu\alpha}F^\mu_{~\,\alpha}-\frac{1}{4}g^{\mu\nu}\left(F_{\alpha_\beta}F^{\alpha_\beta}\right)\right],
\end{equation}
denotes the electromagnetic stress-energy tensor, composed of the electromagnetic field tensor $F^{\mu\nu}$ and the metric tensor $g^{\mu\nu}$. We do not consider the effects of gravitational curvature and assume a Minkowski metric $g^{\mu\nu}\to\eta^{\mu\nu}$ with signature $s=-2$. Combining the energy-momentum equation [Eq.~(\ref{eq:energy-mom})] with the covariant homogeneous and inhomogeneous Maxwell's equations, respectively
\begin{equation}\label{eq:homogen Maxwell}
\partial_\mu (\star F)^{\nu\mu}=0,
\end{equation}
and
\begin{equation}\label{eq:inhomogen Maxwell}
\partial_\mu F^{\mu\nu}=\frac{4\pi}{c} J^\nu,
\end{equation}
where $(\star F)^{\mu\nu}=(1/2)\epsilon^{\mu\nu\sigma\lambda}F_{\sigma\lambda}$ represents the Hodge dual of $F^{\mu\nu}$ and $J^\mu=(c\rho_{\rm ch},\mathbf{j})^T$ is the four-current\footnote{Here $\rho_{\rm ch}$ represents the plasma charge density.}, we may write the divergence of the stress-energy tensor as
\begin{equation}\label{divTf}
\partial_\nu T_{\rm em}^{\mu\nu}-\frac{1}{c}F^{\mu\nu}J_{\nu}=0,
\end{equation}
and subsequently find
\begin{equation}\label{covariantFFEeq}
F^{\mu\nu}J_{\nu}=0.
\end{equation}
The above expression is the so-called force-free condition and implies specifically that the Lorentz force,
\begin{equation}\label{eq}
f^{\mu}=\frac{1}{c}F^{\mu\nu}J_\nu
\end{equation}
is required to vanish, i.e. that the force-free electromagnetic field is fundamentally degenerate \citep{Komissarov02}. It follows immediately that the first electromagnetic invariant is zero,
\begin{equation}\label{eq:1st elec variant}
F_{\mu\nu}(\star F)^{\mu\nu}=0,
\end{equation}
which is known as the degeneracy condition. This means that the inertia of the plasma particles, but not their electromagnetic interaction, is ignored. 

In ideal FFDE we wish to describe the plasma velocity in a physical force-free electromagnetic field. To this end we require the plasma velocity field given by $U^{\mu}=\gamma(c,\mathbf{v})^T$ to satisfy $F_{\mu\nu}U^{\nu} = 0$. Since the four-velocity of the plasma is a time-like vector, we demand that the second electromagnetic invariant is positive,\begin{equation}\label{eq:2nd elec variant}
F_{\mu\nu}F^{\mu\nu} > 0,
\end{equation}
which necessitates the existence of time-like zero eigenvectors of $F_{\mu\nu}$. This condition implies that there exists a reference frame wherein observers at rest detect a field that is purely magnetic, i.e. where the electric field vanishes entirely \citep{Uchida97}.

Adopting 3+1-notation we find that Eq.~(\ref{covariantFFEeq}), Eq.~(\ref{eq:1st elec variant}), and Eq.~(\ref{eq:2nd elec variant}) become respectively
\begin{align}
\rho_{\rm ch}\mathbf{E}+\frac{1}{c}\mathbf{j}\times\mathbf{B}&=0,\label{eq: force-free cond. 3+1D}\\
\mathbf{E}\cdot\mathbf{B}&=0\label{eq:1st elec variant in 3+1D},\\
B^2-E^2&>0\label{eq: 2nd elec variant in 3+1D},
\end{align}
where $E$ denotes the magnitude of the electric field and $B$ the magnitude of the magnetic field. Incidentally, $F^{0i}J_i=\mathbf{E}\cdot \mathbf{j}=0$ and the electromagnetic energy is conserved, i.e. 
\begin{equation}\label{eq: conservation elemag energy}
\partial_t(\mathbf{E}\cdot\mathbf{B})=0.
\end{equation}

To obtain Ohm's law, which describes the relation between the current and the electric field, it is convenient to separate the current vector into components parallel and perpendicular to the magnetic field vector,
\begin{align}\label{jincomponentseq}
\mathbf{j}=\mathbf{j}_\perp+\mathbf{j}_\|,&&\mathbf{j}_\perp=\frac{(\mathbf{B}\times\mathbf{j})\times\mathbf{B}}{B^2},
&&\mathbf{j}_\|=\frac{(\mathbf{B}\cdot\mathbf{j})\,\mathbf{B}}{B^2}.
\end{align}
With the force-free condition Eq.~(\ref{eq: force-free cond. 3+1D}) we may express the perpendicular component as
\begin{equation}\label{eq}
\mathbf{j}_\perp=\rho_{\rm ch}\mathbf{v}_\perp
\end{equation}
where along with the requirement expressed by Eq.~(\ref{eq: 2nd elec variant in 3+1D}) we have defined the electric drift velocity
\begin{equation}\label{eq: drift velocity}
\mathbf{v}_\perp\equiv c\,\frac{\mathbf{E}\times\mathbf{B}}{B^2},
\end{equation}
which denotes the plasma velocity component across the magnetic field. 

In the singular current sheet however the ideal MHD approximation breaks down and the magnetic resistivity becomes finite, i.e. the second electromagnetic invariant [Eq.~(\ref{eq: 2nd elec variant in 3+1D})] becomes negative. Accordingly, the parallel component of Ohm's law is altered to include the effect of current dissipation, solely along the magnetic field, due to the presence of a resistive electric field. To this end we introduce the relativistic formulation of Ohm's law in covariant form \citep{Gedalin96}, 
\begin{equation}\label{eq: rel. Ohms law}
F^{\mu\nu}U_{\nu}=\frac{4\pi}{c}\,\Theta^{\mu\nu}(\delta^\alpha_\nu-U_\nu U^\alpha)J_\alpha,
\end{equation}
where $\Theta^{\mu\nu}$ represents the resistivity tensor. This tensor is highly anisotropic in FFDE, since only the currents flowing along the field may experience resistive dissipation. Accordingly we define the resistivity tensor as such
\begin{equation}\label{eq}
\Theta^{\mu\nu}\equiv\eta\frac{b^\mu b^\nu}{b^2},
\end{equation}
where $b^\mu=(\star F)^{\mu\nu}U_{\nu}$ represents the magnetic four-vector and the scalar resistivity or magnetic diffusivity, which characterises the dissipation of currents, is given by the phenomenological parameter\footnote{We do not derive $\eta$ from microscopic plasma processes, but rather assume a simple macroscopic description.} $\eta=c^2/(4\pi\sigma)$, with $\sigma$ the macroscopic conductivity of the plasma. Subsequently, by convolving Eq.~(\ref{eq: rel. Ohms law}) with the magnetic four-vector we obtain \citep{Lyutikov03}
\begin{equation}\label{eq}
F^{\mu\nu}U_{\nu}(\star F)_{\mu\alpha}U^{\alpha}=\frac{4\pi}{c}\eta(\star F)_{\mu\alpha}U^{\alpha}J^{\mu},
\end{equation}
which in 3+1 notation becomes
\begin{equation}\label{eq}
\gamma^2(\mathbf{B}\cdot\mathbf{E})(c^2-\mathbf{v}\cdot\mathbf{v}) =\frac{4\pi}{c}\eta\gamma\left[\mathbf{j}\cdot(c\,\mathbf{B}-\mathbf{v}\times\mathbf{E})-J^0(\mathbf{B}\cdot\mathbf{v})\right],
\end{equation} 
where $\gamma=[1-(\mathbf{v}\cdot\mathbf{v}/c^2)]^{-1/2}$ is the Lorentz factor. The above expression reduces to
\begin{equation}\label{eq:3+1notation, 2}
\frac{c^2}{4\pi\eta}(\mathbf{B}\cdot\mathbf{E})=\frac{\gamma}{c}\left[\mathbf{j}\cdot(c\,\mathbf{B}-\mathbf{v}\times\mathbf{E})-J^0(\mathbf{B}\cdot\mathbf{v})\right].
\end{equation}
Upon splitting vectors into components parallel and perpendicular to the magnetic field we can rewrite Eq.~(\ref{eq:3+1notation, 2}) as
\begin{equation}\label{eq: parallel EJ}
\frac{c^2}{4\pi\eta}(\mathbf{B}\cdot\mathbf{E})=\gamma \left[(\mathbf{B}\cdot\mathbf{j})\left(1-\frac{E_\perp^2}{B^2}\right)-\rho_{\rm ch}(\mathbf{B}\cdot\mathbf{v})\left(1-\frac{E_\perp^2}{B^2}\right)\right].
\end{equation}
We remove the second term on the r.h.s. by choosing our coordinate system such that $v_\|\equiv0$. Consequently with Eq.~(\ref{eq: drift velocity}) we have that,
\begin{equation}\label{eq}
\gamma^{-2}=\left(1-\frac{E_\perp^2}{B^2}\right),
\end{equation} 
such that Eq.~(\ref{eq: parallel EJ}) becomes
\begin{equation}\label{eq}
(\mathbf{B}\cdot\mathbf{j})=\frac{c}{4\pi}\left[\frac{c\gamma}{\eta}(\mathbf{B}\cdot\mathbf{E})\right].
\end{equation}
Accordingly we use the above result to rewrite the parallel component of the current vector and ultimately obtain the following expression for the current vector,
 \begin{equation}\label{resistiveformofOhmslawRMHDeq}
\mathbf{j}=\frac{c}{4\pi}\left[4\pi \rho_{\rm ch}\frac{\mathbf{v}_\perp}{c}+\frac{c\gamma}{\eta}\frac{(\mathbf{B}\cdot\mathbf{E})\mathbf{B}}{B^2}\right],
\end{equation}
which describes Ohm's law in resistive FFDE, whereby the electric current is written solely in terms of the electric field components, parallel and perpendicular to the magnetic field. Note that in the plasma rest frame $\mathbf{v}_\perp=0$, the electromagnetic field is no longer purely magnetic, due to the presence of the resistive electric field.\\
\\
\subsection{Magnetodynamics near force-free equilibrium}

The divergence of the stress-energy tensor Eq.~(\ref{divTf}) determines the energy- and momentum conservation equations, given in 3+1 notation as follows 
\begin{align}
&\partial_tu_{\rm em}+\boldsymbol{\nabla}\cdot\mathbf{S}+\mathbf{E}\cdot\mathbf{j}=0,\label{eq:3+1 energy}\\
&\partial_t\mathbf{p}_{\rm em}-\boldsymbol{\nabla}\cdot\mathbf{T}^{ij}_{\rm em}+\frac{1}{c}\mathbf{j}\times\mathbf{B}+\rho_{\rm ch} \mathbf{E}=0,\label{eq:3+1 moment}
\end{align} 
where respectively
\begin{equation}\label{eq}
u_{\rm em}=\frac{B^2+E^2}{8\pi}~~~~\text{and}~~~~\mathbf{p}_{\rm em}=\frac{\mathbf{S}}{c^2},
\end{equation}
are the electromagnetic energy- and electromagnetic momentum density. The above expressions are written in terms of the Poynting vector,
\begin{equation}\label{eq}
\mathbf{S}=\frac{c}{4\pi}\mathbf{E}\times\mathbf{B},
\end{equation}
and the Maxwell stress-tensor
\begin{equation}\label{eq}
\mathbf{T}^{ij}_{\rm em}=\frac{1}{4\pi}\left[E^iE^j+B^iB^j-\frac{1}{2}(E^2+B^2)\delta^{ij}\right],
\end{equation}
where $\delta^{ij}$ is the Euclidean metric of flat space.

To study the dynamical properties of a system near force-free equilibrium, we introduce the relevant timescales via the relativistic Lundquist number,
\begin{equation}\label{eq:Lundqvist number}
\mathcal{S}_l\equiv\frac{\tau_\eta}{\tau_A}=\frac{l c}{\eta},
\end{equation}
where $\tau_\eta\equiv l^2/\eta$ is the resistive diffusion timescale, $\tau_A\equiv l/v_A\to l/c$ denotes the hydromagnetic timescale or Alfv\'en transit time (for $\sigma_{\rm m}\gg1$), and $l$ denotes the corresponding typical length scale of the system. 

The evolution of the system can be represented by the timescale $\tau$, for which $\tau_A\ll\tau\ll\tau_\eta$. Accordingly, $|\mathbf{v}_\perp|\ll c$ and with Eq. (\ref{eq: drift velocity}) we find naturally $E_\perp\ll B$. Immediately we may approximate,
\begin{align}
&\gamma\to1,\nonumber\\
&u_{\rm em}\simeq u_B=\frac{B^2}{8\pi},\nonumber\\
&\mathbf{T}^{ij}_{\rm em}\simeq\frac{1}{4\pi}\left(B^iB^j-\frac{B^2}{2}\delta^{ij}\right).\nonumber
\end{align}

Scaling Eq. (\ref{eq:3+1 energy}) and Eq. (\ref{eq:3+1 moment}) in terms of the small expansion parameters ($\tau/\tau_\eta$) and ($\tau_A/\tau$) and assuming incompressibility of the plasma, \citet{Komissarov07} derive the following closed set of equations,
\begin{align}
&\boldsymbol{\nabla}\cdot\mathbf{v_\perp}=0,\label{eq:system MD1}\\
&\boldsymbol{\nabla}\cdot\mathbf{B}=0,\\
&\partial_t \mathbf{B}=\boldsymbol{\nabla}\times(\mathbf{v_\perp\times B})+\eta\nabla^2 \mathbf{B},\label{eq:system MD3}\\
&\rho_{\rm em}[\partial_t(\boldsymbol{\nabla}\times\mathbf{v}_\perp)]=\frac{1}{8\pi}\boldsymbol{\nabla}\times(\mathbf{B}\cdot\boldsymbol{\nabla})\mathbf{B},\label{eq:system MD4}
\end{align}
that together govern the dynamics of a system near force-free equilibrium and incidentally closely resemble the equations of non-relativistic resistive incompressible MHD.

\subsection{Growth time of the (relativistic) tearing mode}

\subsubsection{Linear stability analysis}

\begin{figure}
	\centering
		\includegraphics[width=0.475 \textwidth]{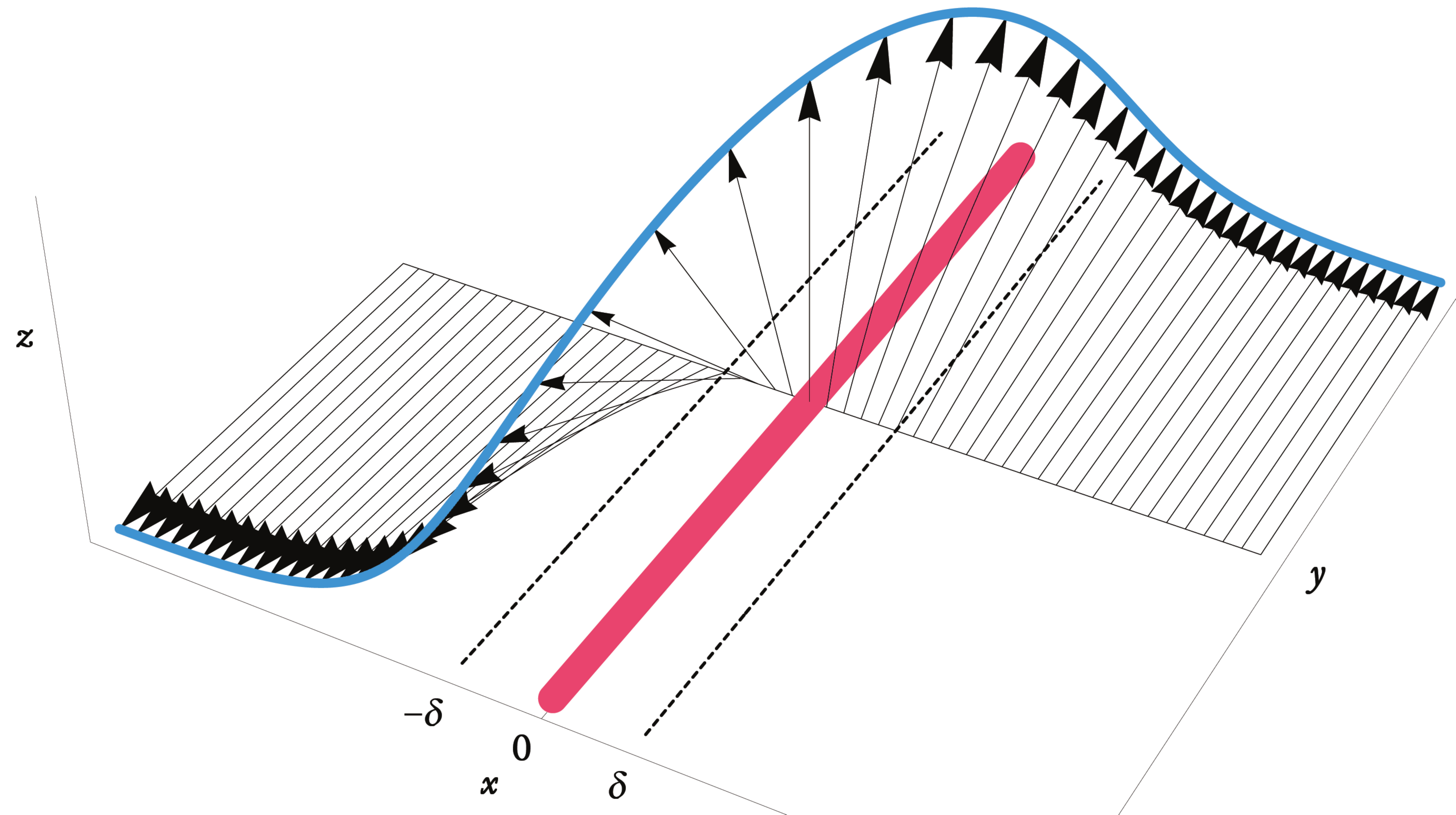}
	\caption{3D schematic representation of the force-free equilibrium profile in the form of a rotational discontinuity. The magnitude of the magnetic field vector remains constant under rotation over $\pi$ radian and most of the rotation takes place within $-\delta < x < \delta$. Accordingly, the typical length scale of the system, i.e. the current sheet (half-)thickness, is given by $\delta$. A magnetic null line, which denotes the location of the resistive sublayer, is formed in the $xy$-plane at $x = 0$. }\label{fig:equi2}
\end{figure}

The growth time of the tearing instability may be obtained by performing linear stability analysis on a current sheet described by the following one-dimensional force-free equilibrium profile that represents a rotational discontinuity \citep{Low73},
\begin{equation}\label{eq:}
\mathbf{B}_0 = B_0\tanh\left(\frac{x}{\delta}\right) \hat{\mathbf{y}}  \pm B_0\,{\rm sech}\left(\frac{x}{\delta}\right) \hat{\mathbf{z}},
\end{equation}
where the magnetic null line is given by the sheared $B_{0y}$-component that goes to zero at $x=0$, whilst the magnitude of the magnetic field vector $|\mathbf{B}_0(x)|$ remains constant under rotation over $\pi$ radian (see Figure~\ref{fig:equi2}). The vector rotates predominantly within the domain $-\delta<x<\delta$, such that the typical length scale of the system is given by $l\to\delta$, which denotes the (half-)thickness of the current sheet\footnote{In the following we refer to $\delta$ simply as the thickness of the current sheet, even though in principle it only describes half of the total thickness - see Figure \ref{fig: mag_sphere1}.}.

Through linearising the dynamical equations of resistive MD\footnote{The linearised equations of resistive MD are equal to those of resistive MHD, such that the growth time of the linear tearing mode remains equal for both regimes. This similarity was first made explicit by \citet{Komissarov07}.} \citet{Komissarov07} demonstrate how to derive the following expressions for the (maximum) wavelength and (minimum) growth time of the fastest growing linear mode,
\begin{align}\label{eq: tearing mode timescale (general)}
&\lambda^{\rm max}=2\pi\,\delta\mathcal{S}_\delta^{1/4},
&\tau_{\rm tm}^{\rm min}=\tau_A\mathcal{S}_\delta^{1/2}=\left(\tau_A\tau_\eta\right)^{1/2},
\end{align}
where $\mathcal{S}_\delta=c\delta/\eta$ is the relativistic Lundquist number corresponding to the length scale $\delta$ and the minimum growth time of the tearing mode is ascertained to be the geometric mean of the Alfv\'en- and resistive diffusion timescale, as in the case of non-relativistic resistive incompressible MHD. Further comprehensive and general derivations of tearing mode characteristics may be found in the literature, e.g. \citet{White86}, \citet{Goldston95}, \citet{Priest00}, \citet{Lyutikov03}, and \citet{Goedbloed09}.

\subsection{Tearing mode growth time in magnetar magnetospheres}

Here we aim to establish the minimum growth time of the tearing mode prevailing in magnetar magnetospheres. In a globally twisted magnetic field the radial dependency of the magnetic field strength is approximately given by \citep{Thompson02}
\begin{equation}\label{eq:mag. r dependence}
B_0(r)\simeq B_{\rm s}\left(\frac{r}{R_*}\right)^{-(2+p)},
\end{equation}
where $B_{\rm s}$ denotes the inferred surface dipole magnetic field strength and $R_*\sim10^6$ cm is the typical NS radius. Also, $0<p<1$ is the radial index which parameterises the net twist angle $0<\Delta\phi<\pi$, where the limiting value $p=1$ ($p=0$) corresponds to a net twist of $\Delta\phi=0$ ($\Delta\phi=\pi$), representing a pure dipole (split monopole) configuration. $B_0(r)$ will function as the background or upstream magnetic field strength of the reconnection region.

We need a qualitative estimate of the local magnetic resistivity $\eta$. We consider a macroscopic description, whereby the resistivity is homogeneous and given by the presence of Langmuir turbulence \citep[as in][]{Lyutikov03}. In this case the typical turbulent length scale is given by the electron skin depth,
\begin{equation}\label{eq}
\delta_e=\frac{c}{\omega_{p,e}}
\end{equation} 
where $\omega_{p,e}=(4\pi n_\pm e^2/m_e)^{1/2}$ is the electron plasma frequency, with $n_\pm=n_++n_-$ the total number density of the charge carriers, i.e. the sum of positrons $n_+$ and electrons $n_-$, $e$ is the elementary charge unit, and $m_e$ the electron mass. Accordingly, the resultant resistivity of a turbulent plasma with a typical eddy size and turnover velocity of $\delta_e$ and $c$ respectively, is approximately
\begin{equation}\label{eq: resistivity ito plasma freq}
\eta\sim c\,\delta_e=\frac{c^2}{\omega_{p,e}}=c^2\left(\frac{4\pi n_\pm e^2}{m_e}\right)^{-1/2}.
\end{equation}
In the aforementioned globally twisted dipole model the magnetospheric currents generate a toroidal field component that approaches the strength of the poloidal field, i.e. $B_t\lesssim B_p\sim B_{0}(r)$. Therefore we may apply Amp\`ere's law to obtain an estimate for the charge number density as a function of the local magnetic field strength $B_0(r)$ and distance from the NS centre $r$ \citep{Lyutikov02},
\begin{equation}\label{eq: Ampere to plasma freq.}
\boldsymbol{\nabla}\times\mathbf{B}_0=\frac{4\pi}{c}\mathbf{j}=4\pi e\,[\boldsymbol{\beta}_+n_+-\boldsymbol{\beta}_-n_-],
\end{equation}
where $\boldsymbol{\beta}_+$ and $\boldsymbol{\beta}_-$ are the dimensionless drift velocities of the positrons and electrons, respectively.
If we consider the case where $\boldsymbol{\beta}_+=-\boldsymbol{\beta}_-\sim\mathbf{1}$ and $n_+\simeq n_-$, we may simplify\footnote{Twisted magnetospheres are believed to be threaded by pairs moving at mildly relativistic speeds and with low multiplicity, as required to explain  magnetars quiescent emission at X-ray energies \cite[see e.g.][and references therein]{Turolla15}.}
\begin{equation}\label{eq:}
n_\pm\sim\frac{B_0(r)}{8\pi e\,r}.
\end{equation}
Accordingly, we obtain an expression for the plasma frequency,
\begin{equation}\label{eq:}
\omega_{p,e}\sim\left(\frac{\omega_{c,e}\,c}{r}\right)^{1/2},
\end{equation}
with the electron cyclotron frequency given by $\omega_{c,e}\equiv eB_0(r)/(m_e c)$. The resistivity as a function of the surface dipole magnetic field strength and distance to the centre of the NS becomes
\begin{equation}\label{eq: resistivity}
\eta\simeq c^2\left[\frac{eB_{\rm s}}{m_er}\left(\frac{r}{R_*}\right)^{-(2+p)}\right]^{-1/2}.
\end{equation}
Now together with Eq.~(\ref{eq: tearing mode timescale (general)}) and Eq.~(\ref{eq:Lundqvist number}) we may ultimately obtain the minimum growth time of the tearing mode in magnetar magnetospheres,
\begin{align}\label{eq: tearing mode timescale (general+resistivity)}
&\tau_{\rm tm}^{\rm min}=\left(\frac{\delta^3}{c\eta}\right)^{1/2}\simeq\left(\frac{eR_*^{(2+p)}}{m_e c^6}\right)^{1/4}\,\delta^{3/2}r^{-(3+p)/4}B_{\rm s}^{1/4}.
\end{align}
In order to compare this result with the observed timescales, we rewrite the above result in terms of typical values for the relevant parameters,
\begin{equation}\label{eq:minimum growth time linear tearing mode}
\tau_{\rm tm}^{\rm min}\simeq10^{-1}\,\delta_4^{3/2}r_7^{-(3+p)/4}B_{\rm s, 15}^{1/4}~{\rm ms},
\end{equation}
where we define $\delta_4\equiv\delta/(10^4$ cm), $r_7\equiv r/(10^7$ cm), $B_{\rm s, 15}\equiv B_{\rm s}/(10^{15}$ G), and $0< p< 1$ (in practice $p$ will always be close to unity). With these scalings, the minimum growth time of the tearing mode agrees nicely with the observed (sub)millisecond $e$-folding rise times $\tau_e$ of the magnetar giant flares. 

Note that this timescale differs significantly from the minimum growth time as calculated by \citet{Lyutikov03}, essentially due to an error in that calculation (specifically in the inferred expression for the plasma frequency). In addition we have adopted a rather smaller (by a factor of $10^{-2}$) typical size for the thickness of the current sheet $\delta$ than the value used in \citet{Lyutikov03}. We do this since for large gradients to develop, the thickness of the current sheet must be significantly less than the global extent of the reconnection region, which in the case of magnetar giant flares is a few times the NS radius. \citet{Komissarov07} argue for a current sheet thickness of $\sim3\times10^{3}$ cm, however we have not been able to reproduce their inferred tearing mode timescale (particularly Eq. (73) in their paper).   Without the above modification to the typical value for $\delta$, however, the inferred tearing mode growth time would be $\sim 100$ ms \citep{Duncan04}.  If this were the case, it would entirely rule out the development of the tearing mode as a candidate mechanism to explain the (sub)millisecond rise times of the magnetar giant flares.

In the subsequent section we will assume that the trigger is given by the development of a tearing instability, and that its minimum growth time corresponds to the timescale on which the observed emission is released from the system, i.e. $\tau^{\rm min}_{\rm tm}=\tau_e$.   We explore additional constraints on the geometry of the reconnection region that are required for the linear tearing mode to be a plausible mechanism for the giant flares, and discuss how they relate to the constraints derived in this section.  

\section{Physical constraints on the reconnection region}\label{sec:pcotrr}

Here we present two straightforward models, respectively based on energy conservation and mechanical equilibrium of the current sheet, that provide order of magnitude estimates for the  thickness of the current sheet $\delta$ and height of the base of the reconnection region in terms of the radial distance from the NS centre $r$. 

In order to relate the thickness of the tearing unstable current sheet to the global length of the reconnection region $L_y$, we consider the following elementary instability condition: for an unstable mode to be able to develop in a current sheet, its growth time ($\tau_{\rm tm}^{\rm min}$) is required to be less than the time it would take for the perturbation to exit the system ($L_y/c$) \citep{Shibata01}. We obtain the requirement
\begin{equation}\label{eq: cond1}
\tau_{\rm tm}^{\rm min}<\frac{L_y}{c}.
\end{equation}
Together with Eq.~(\ref{eq: tearing mode timescale (general+resistivity)}) we have the following upper limit to the thickness of the current sheet
\begin{equation}\label{eq: cond2}
\delta^{\rm max}=\mathcal{S}_\delta^{-1/2} L_y,
\end{equation}
and equivalently
\begin{equation}\label{eq: cond3}
\delta^{\rm max}=[\,c\,\eta\,(\tau^{\rm min}_{\rm tm})^2\,]^{1/3}.
\end{equation}

\subsection{Conversion of magnetic energy}\label{subsec:ec}

\begin{figure}
	\centering
		\includegraphics[width=0.475 \textwidth]{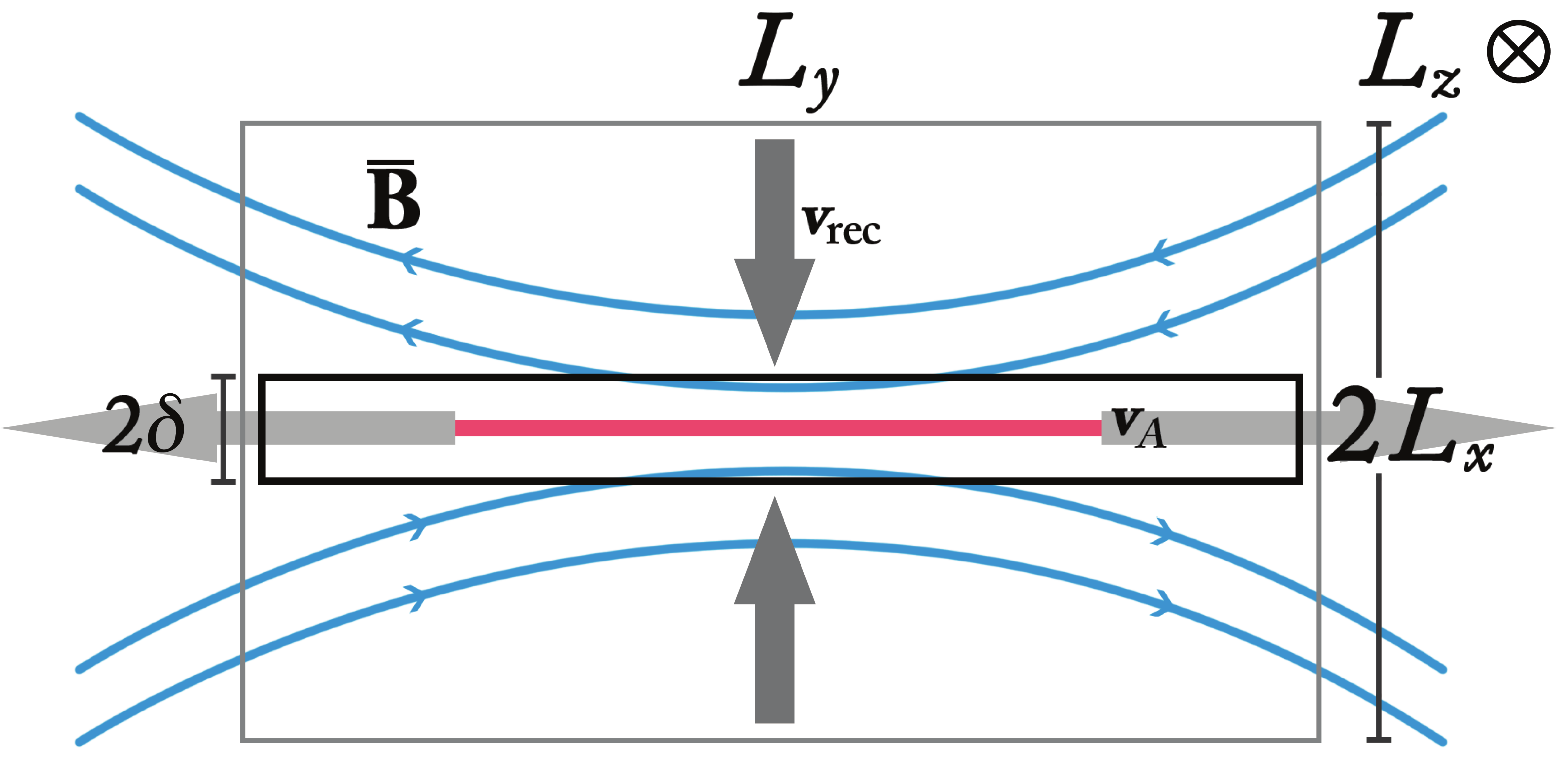}
	\caption{2D schematic representation of the reconnection region -- the reconnection geometry is uniform along the $z$-direction. The curved (blue) arrows denote the sheared magnetic field component, the thick arrows (grey) arrows represent the plasma in- and outflows, the large box describes the extent of the reconnection region for the duration of the hard $\gamma$-ray spike, and the smaller rectangular box denotes the current sheet. The volume of the entire reconnection region is given by $V=(2L_x)L_yL_z$, where $L_x=v_{\rm rec}\tau_{\rm spike}=(\delta/\tau_{\rm rec})\tau_{\rm spike}$, representing the extent to which the magnetic field is fed into the diffusion region for the duration of the spike, assuming $v_{\rm rec}$ remains constant, and $L_z<2\pi r$ for an equatorial current sheet. This image essentially represents a magnification of the reconnection region depicted in Figure~\ref{fig: mag_sphere1}.}\label{fig:rec1}
\end{figure}

Figure~\ref{fig:rec1} shows the geometry of the reconnection region in the $xy$-plane, whereby the curved (blue) arrows represent the sheared magnetic field that continues to annihilate within the current sheet, which in turn is denoted by the smaller rectangular box ($2\delta\times L_y$). The larger rectangle ($2L_x\times L_y$) describes the size of the total area that proceeds to reconnect, i.e. the extent of magnetic flux that is advected into the current sheet for the duration of the hard $\gamma$-ray spike $\tau_{\rm spike}$. We hypothesise that the magnetic field lines are fed into the diffusion region at a constant rate (this assumption is discussed further in section \ref{ssec:tearing mode phases}). The reconnection rate is generally determined by the aspect ratio of the reconnection region through mass flux conservation \citep[e.g.][]{Pucci13}, i.e
\begin{equation}\label{eq:}
\frac{v_{\rm rec}}{c}\simeq\frac{\delta}{L_y}=\mathcal{S}_\delta^{-1/2},
\end{equation}
which together with Eq. (\ref{eq: cond1}) leads to
\begin{equation}\label{eq:}
v_{\rm rec}\simeq\frac{\delta}{\tau_{\rm rec}}.
\end{equation}
Accordingly, we find that
\begin{equation}\label{eq: Lx approx}
L_x\sim v_{\rm rec}\tau_{\rm spike}=\delta\left(\frac{\tau_{\rm spike}}{\tau_{\rm rec}}\right).
\end{equation}
For an equatorial current sheet we have $L_z < 2\pi r$ (one may picture the current sheet as a disk around the NS if $L_z=2\pi r$). The entire volume of magnetic flux that reconnects over the course of the initial hard phase of the giant flare is then given by $V=(2L_x)L_yL_z$. The energy contained in this region, that is subsequently released within $\tau_{\rm spike}$ can be estimated as
\begin{equation}\label{eq}
E_{\rm tot}\simeq \zeta\, u_BV=\zeta\,\frac{B_0^2}{8\pi}\left(2L_xL_yL_z\right)=\frac{\zeta\,B_0^2r L_y \delta}{2} \left(\frac{\tau_{\rm spike}}{\tau_{\rm rec}}\right),
\end{equation}
where $\zeta$ is the fraction of free magnetic energy that is dissipated and we have used $u_B=B_0^2/(8\pi)$ for the local magnetic energy density in terms of the upstream magnetic field $B_0$. Rewriting this equation we obtain 
\begin{equation}\label{eq}
\delta(r)\sim \frac{2\,E_{\rm tot}}{\zeta\,B_0^2rL_y}\left(\frac{\tau_{\rm rec}}{\tau_{\rm spike}}\right).
\end{equation}
Note incidentally that the above general expression does not rely on any particular reconnection mechanism as yet. If we now consider linear tearing as the principal reconnection mechanism, we may set $\tau_{\rm rec}=\tau_{\rm tm}^{\rm min}$ and, through Eq. (\ref{eq: cond1}), $L_y= c\tau_{\rm tm}^{\rm min}$. Using Eq. (\ref{eq:mag. r dependence}) and adopting $p=1/2$ we end up with
\begin{equation}\label{eq}
\delta(r)\sim\frac{2\,E_{\rm tot}r^4}{\zeta\,cB_s^2R_*^5\tau_{\rm spike}}.
\end{equation}
Together with the condition stated in Eq. (\ref{eq: cond3}), we find an estimate for the height of the reconnection region
\begin{equation}\label{eq:e conserv r}
r_{\rm rec}\sim 10^7\left[\zeta\,B_{s,15}^{11/6} (\tau_{\rm tm,-4}^{\rm min})^{2/3} \left(\frac{\tau_{\rm spike,0.2}}{E_{\rm tot,45}}\right)\right]^{12/41}~{\rm cm},
\end{equation}
and the thickness of the current sheet at $r_{\rm rec}$,
\begin{equation}\label{eq:e conserv delta}
\delta(r_{\rm rec})\sim10^4\left[\zeta^7\,B_{s,15}^{6} (\tau_{\rm tm,-4}^{\rm min})^{32} \left(\frac{\tau_{\rm spike,0.2}}{E_{\rm tot,45}}\right)^{7}\right]^{1/41}~{\rm cm}.
\end{equation}
In the above we have made use of Eq. (\ref{eq: resistivity}) and Eq. (\ref{eq: cond1}) to eliminate $\eta(r)$ and $L_y$ respectively. The solutions depend mildly on $\zeta$.

\subsection{Mechanical equilibrium}

Without mechanical equilibrium across the current sheet boundary, the current sheet would disrupt before reconnection could occur effectively \citep{Uzdensky11}. This requirement is given by the following pressure balance,
\begin{equation}\label{eq}
P_{\rm cs}+\frac{B_{\rm cs}^2}{8\pi}=P_0+\frac{B_0^2}{8\pi},
\end{equation}
where $P_{\rm cs}$ and $B_{\rm cs}$ respectively are the leptophotonic pressure [see Eq. (\ref{eq:leptophotonic pressure})] and magnetic field strength inside the current sheet, and $P_0$ and $B_0$ respectively are the local plasma pressure and magnetic field strength in the upstream region. In the upstream region we have $\sigma_{\rm m}\gg1$, such that the plasma beta, $\beta=P_{\rm plasma}/P_{\rm mag}$, is small, i.e. $P_0\ll B_0^2$. Consequently, the above expression simplifies to
\begin{equation}\label{eq: reduced balance}
P_{\rm cs}+\frac{B_{\rm cs}^2}{8\pi}\simeq \frac{B_0^2}{8\pi}.
\end{equation}
The leptophotonic pressure in the current sheet may be decomposed as
\begin{equation}\label{eq:leptophotonic pressure}
P_{\rm cs}=P_{\rm rad}+P_{\pm},
\end{equation}
where $P_{\rm rad}$ signifies the radiation pressure and $P_{\pm}$ denotes the pressure as a result of pair production. In a relativistic current sheet $P_{\pm}$ becomes $\sim (7/4)P_{\rm rad}$ \citep{Uzdensky11}, such that
\begin{equation}\label{eq: Pcs is fPrad}
P_\text{cs}\sim \frac{11}{4} P_{\rm rad},
\end{equation}
and
\begin{equation}\label{eq}
P_{\rm rad}(T_{\rm cs})=\frac{4\,\sigma_{\rm SB}}{3\,c\,k_{\rm B}^4}(k_BT_{\rm cs})^4,
\end{equation}
where $\sigma_{\rm SB}\equiv\pi^2k_{\rm B}^4/(60\hbar^3c^2)\simeq5.67\times10^{-5}$ erg cm$^{-2}$ s$^{-1}$ K$^{-4}$ is the Stefan-Boltzmann constant and $T_{\rm cs}$ represents the temperature inside the current sheet.

Eq. (\ref{eq: reduced balance}) may then be written as
\begin{equation}\label{eq: reduced balance2}
B_0^2-B_{\rm cs}^2=22\pi P_{\rm rad}.
\end{equation}
Furthermore using Gauss's law for magnetism $\boldsymbol{\nabla}\cdot\mathbf{B}=0$, we approximate
\begin{equation}\label{eq}
\frac{B_{\rm cs}}{\delta}+\frac{B_y}{L_y}+\frac{B_g}{L_z}\simeq0,
\end{equation}
where we respectively parameterise the strengths of the guide field and $y$-component of the field as $B_g=qB_0$ and $B_y=(1-q)B_0$, with $0\leq q \leq1/2$ and $\mathbf{B}_0=\mathbf{B}_g+\mathbf{B}_y$.
Subsequently, we may write
\begin{equation}\label{eq:}
B_{\rm cs}\simeq B_0 \left[(1-q)\frac{\delta}{L_y}+q\frac{\delta}{L_z}\right].
\end{equation}
Together with Eq. (\ref{eq: cond2}) and the relation for $L_z$ below Eq. (\ref{eq: Lx approx}) this becomes
\begin{equation}\label{eq: Bcs as function of B0}
B_{\rm cs}\simeq B_0 \left[(1-q)\,\mathcal{S}_\delta^{-1/2}+q\frac{\delta}{2\pi r}\right],
\end{equation}
such that we may eliminate $B_{\rm cs}$ from Eq. (\ref{eq: reduced balance2}): 
\begin{equation}\label{eq: reduced balance3}
B_0^2\left\{1- \left[(1-q)\,\mathcal{S}_\delta^{-1/2}+q\frac{\delta}{2\pi r}\right]^2\right\}=22\pi P_{\rm rad}.
\end{equation}
The above equation depends on the values for $B_s$, $T_{\rm cs}$, $\delta$ and $r$. To solve Eq.~(\ref{eq: reduced balance3}) we need to write $\delta$ in terms of $r$ via Eq. (\ref{eq: cond3}) and require an estimate for the temperature \emph{inside} the current sheet $k_{\rm B}T_{\rm cs}$. It is however questionable whether $k_{\rm B}T_{\rm spec}$ -- listed in Table \ref{tab:gf data} -- would represent $k_{\rm B}T_{\rm cs}$, since the former may rather correspond to a Lorentz boosted photospheric temperature of a relativistically expanding fireball. Of necessity, we consider here the following reasonable range of temperatures: $k_{\rm B}T_{\rm cs}\sim250-1000$ keV.

Consequently, together with $B_s=10^{15}$ G and $\tau_{\rm tm}^{\rm min}=10^{-4}$ s, we solve Eq. (\ref{eq: reduced balance3}) numerically for $r$ and find,
\begin{equation}\label{eq}
r_{\rm rec}\sim(3\times10^6)-10^{7}~{\rm cm},
\end{equation}
and furthermore with Eq.~(\ref{eq: cond3}) we have
\begin{equation}\label{eq}
\delta(r_{\rm rec})\sim(4-8)\times10^3~{\rm cm},
\end{equation}
where the lower (upper) estimates of the above equations correspond to the upper (lower) value for $k_{\rm B}T_{\rm cs}$. These estimates remain equal down to the fourth decimal for the entire range of $q$ and as one can observe from Eq. (\ref{eq: Bcs as function of B0}), $B_{\rm cs}\ll B_{0}$, such that the second term on the l.h.s. of Eq. (\ref{eq: reduced balance2}) may be neglected to find the following expression (for $p=1/2$),
\begin{equation}\label{eq:}
r_{\rm rec}\sim10^7B_{s,15}^{2/5}\left(\frac{k_{\rm B}T_{\rm cs}}{250\,\text{keV}}\right)^{-4/5}~{\rm cm}.
\end{equation}
Note that the results agree roughly with those obtained in the previous section. Additionally, we find that the dimensionless reconnection rate is approximately $M_{\rm rec}\equiv\delta/(v_A\tau_{\rm rec})\simeq\delta/(c\tau_{\rm tm}^{\rm min})\sim10^{-3}$, which is comparable to the reconnection rates found for solar flares \citep[e.g][]{Narukage06}. Moreover, note that the reconnection region is located high up in the magnetosphere, such that the background magnetic field is sub-critical $B_0(r_{\rm rec})\simeq10^{12}\text{ G}\sim10^{-1}B_{\rm qed}$.

\section{Discussion}\label{sec:ep}

\subsection{Geometry of the reconnection region}

The previous calculations provide estimates for the scale of the reconnection region involving spontaneous tearing of a global current sheet; the sheet length [from Eq.~(\ref{eq: cond1})] is $L_y\gtrsim c\,\tau_{\rm tm}^{\rm min}=c\,\tau_e\sim(3\times10^6)-10^7$ cm, the sheet thickness is $\delta\sim10^4$ cm, and the height of the reconnection region is $r\sim10^7$ cm. Here we briefly discuss various consequences of these results.

We have assumed that the resistivity is given by a homogeneous background of Langmuir turbulence, which fundamentally requires that the drift velocity of the current-carrying particles exceeds the thermal velocity of the background plasma. This needs to be the case throughout the extensive reconnection region ($>2\delta\times L_y$) for impulsive tearing to be able to occur on the requisite short timescales. 
 
With an estimate for the thickness of the reconnection region, we can infer the temperature at the photosphere of the current sheet $k_{\rm B}T_{\rm phot}$ \citep{Uzdensky11}. At the photosphere the optical depth
\begin{equation}\label{eq:optical depth}
\tau\sim\frac{\delta}{\lambda_{\rm mfp}},
\end{equation}
will be of order unity, where $\lambda_{\rm mfp}$ is the photon mean free path. Assuming that this temperature is sub-relativistic, such that the pair number density is given by
\begin{equation}\label{eq:npairs}
n_{\pm}=\frac{1}{\sqrt{2\pi^3}}\left(\frac{m_ec}{\hbar}\right)^{3}\left(\frac{k_{\rm B}T_{\rm phot}}{m_e c^2}\right)^{3/2}\exp\left[-\frac{m_e c^2}{k_{\rm B}T_{\rm phot}}\right],
\end{equation}
and considering that the scattering opacity of O-mode (i.e. ordinary mode) photons in the presence of a strong magnetic field remains close to Thompson scattering opacity, $\sigma_{\rm es}($O$)\sim\sigma_T\equiv(8\pi/3)\,e^4/(m_e c^2)^2\simeq6.65\times10^{-25}$ cm$^2$, we have
\begin{equation}\label{eq:}
\lambda_{\rm mfp}(\text{O})\sim\frac{1}{n_{\pm}\,\sigma_T}.
\end{equation}  
Together with Eq.~(\ref{eq:optical depth}) we find
\begin{equation}\label{eq:delta to Tphot}
\delta\,\sigma_T\,n_{\pm}(k_{\rm B}T_{\rm phot})\sim1,
\end{equation}
which can be solved for $\delta\sim10^4$ cm to get $k_{\rm B}T_{\rm phot}\sim27$ keV. Note however that $k_{\rm B}T_{\rm phot}$ depends only weakly on $\delta$. 

Due to the release of high-energy radiation following the reconnection process, extensive pair-production has resulted in a high photospheric pair density $n_\pm(k_{\rm B}T_{\rm phot}\sim27 \text{ keV})\sim10^{20}$ cm$^{-3}$. Note that this pair density greatly exceeds the charge density that is available prior to the onset of reconnection [from Eq. (\ref{eq: Ampere to plasma freq.}) we establish $n\gtrsim10^{14}B_{\rm s,\,15}r_7^{-7/2}$ cm$^{-3}$]. It is argued that the observed spectral temperatures $k_{\rm B}T_{\rm spec}$ (see Table \ref{tab:gf data}) correspond to a Lorentz-boosted photospheric temperature of a pair fireball that, in the wake of the onset of the flare, expands outward from the stationary reconnection region relativistically \citep{Lyutikov06,Uzdensky11}, such that
\begin{equation}\label{eq:}
\Gamma\,k_{\rm B}T_{\rm phot}=k_{\rm B}T_{\rm spec},
\end{equation}
where $\Gamma$ denotes the bulk Lorentz factor of the ejected fireball.\footnote{Note that the photosphere of the relativistically expanding fireball differs from the stationary emission region associated with the onset of the flare, such that the bulk Lorentz factor of the latter is zero.} Using the result from Eq.~(\ref{eq:delta to Tphot}) and assuming that the dimensions of the fireball roughly correspond to those of the initial current sheet, we obtain $\Gamma\sim10$, which is consistent with previous estimates in literature.

Furthermore, considering the required scale of the initial configuration $L_y$, uniquely determined by $\tau_e$, spontaneous tearing seems an unlikely candidate for the smaller recurrent $\gamma$-ray bursts ($\lesssim10^{41}$ erg, $\tau_e\sim1$ ms \citep{Gogus01,Gavriil04}), since their $e$-folding rise times are similar to those of the giant flares, such that $L_y\sim(3-10)R_{*}$. These particular bursts may rather demonstrate for instance driven reconnection through an external driver (e.g. sudden crustal motion at magnetic foot points or ideal instabilities in smaller critically sheared magnetic arcades (Browning et al. 2008)), or comprise explosive seismic events without involving magnetospheric reconnection altogether. 

\subsection{Linear tearing and the observed high energy emission}

In discussing linear tearing as a candidate mechanism for explaining the fast initial rise of magnetar giant flare light curves, it has been implicitly assumed throughout the literature that the growth of the resistive instability directly coincides with the conversion of magnetic energy -- via Ohmic heating and particle acceleration -- into the observed high energy radiation (i.e. $\tau_{\rm tm}^{\rm min}=\tau_e$) \citep{Lyutikov03,Duncan04,Komissarov07}. This conjecture presumes that (i) linear tearing dictates the rate of radiation release and (ii) that during the linear tearing phase a significant amount of magnetic energy is converted efficiently to produce the observed radiation in the first place. Both assumptions will be examined further; in Section \ref{ssec:nonthermal emission} we discuss the former requisite (i), and in Sections \ref{ssec:tearing mode phases} and \ref{ssec:coalescence instability} we address the latter (ii).

\subsubsection{Nonthermal emission from accelerated particles}\label{ssec:nonthermal emission}

Concerning point (i) above it should be emphasised by observing that even for comparatively well-studied phenomena like solar flares the generation and release of radiation is not unequivocally linked to the reconnection rate. Note that whilst solar flares are not supposed to be directly analogous, the comparison may still be informative. The rapid onset of a solar flare is given by the sudden increase of hard X-ray (HXR) emission due to collisional thick-target Bremsstrahlung interactions of nonthermal particles at the chromospheric footpoints of coronal loop structures undergoing magnetic reconnection \citep{Shibata11}. Accordingly, the observed radiation timescales are determined by the acceleration timescales of the nonthermal particles. 

Proposed acceleration mechanisms include direct acceleration by reconnection induced or field-aligned electric fields \citep[e.g.][]{Aschwanden06,Egedal12}, acceleration through shocks \citep{Aschwanden06}, and stochastic acceleration through turbulence excited by reconnection outflows at the loop top or cascading Alfv\'en waves near the footpoints \citep[e.g.][]{Petrosian04,Liu08,Liu09,Fletcher08}. None of the above processes guarantee a straightforward connection between the timescales of linear tearing and that of radiation release. Moreover, such acceleration mechanisms generally rely on the later phases of reconnection (e.g. nonlinear tearing; see section \ref{ssec:coalescence instability}) or rather its large-scale effects, such as reconnection jets that excite MHD turbulence or the catastrophic rearrangement of the global magnetic field topology. In the latter case, the amount and rate of energy release will be determined more by the dynamic restructuring of the field, than on the dissipation of an extended current sheet \citep{Hoshino12}. 

Efficient particle acceleration in magnetar magnetospheres may however require local regions where the magnetic field becomes small enough, since considerable synchrotron losses might otherwise impede any significant acceleration. Acceleration through reconnection induced electric fields localised at magnetic x-points seems fitting in this regard, since not only does $B_y\to0$ but the presence of $\mathbf{E}\times\mathbf{B}$-drift also focusses the trajectory of the charged particles in the acceleration region \citep{Speiser65}. Particle-in-cell (PIC) simulations of relativistic reconnection in pair plasmas disclose short acceleration timescales \citep{Zenitani01}, such that the timescale on which the radiation is generated is the reconnection rate.

Nonetheless, a major complication is given by the copious pair production that will ensue upon release of high energy radiation in the presence of an ultra-strong magnetic field \citep[e.g.][]{Harding06}, causing the reconnection region to become optically thick. The observed radiation timescales will therefore not necessarily represent the timescales associated with reconnection dynamics \citep{Uzdensky11,Hoshino12}. To further constrain magnetar burst trigger mechanisms, via emission timescales, will require a better understanding of radiation transport in the magnetar magnetosphere.

\subsubsection{Phases of tearing: linear and nonlinear}\label{ssec:tearing mode phases}

Exponential growth of the magnetic island proceeds until their half-width
\begin{equation}\label{eq:}
w(t)\propto\exp\left[\frac{t}{\tau_{\rm tm}}\right],
\end{equation}
becomes comparable to the size of the resistive sublayer $\epsilon\delta$; here nonlinear effects become important. Analytic calculations disclose a transition from exponential to algebraic growth ($\propto t^\alpha$), once this stage is reached \citep{Rutherford73}. Numerical simulations confirm this strong change in reconnection rate, even though it is less significant when $k\ll1$ and $\mathcal{S}_\delta\gg1$ \citep{Steinolfson84}. Moreover, it is found that the nonlinear regime sets in very quickly, after only a few $e$-folding times, such that one would expect to observe a considerable change in reconnection rate just moments after the onset of the instability. With $\tau_{\rm tm}^{\rm min}\sim10^{-4}$ ms, the exponential phase of the light curve would only last a few tenths of milliseconds to a millisecond, followed by a notable decline in count rate due to the transition from linear to nonlinear tearing. A break in the increase of the count rate during the initial rise to peak has been observed for the SGR 1806-20 flare by \citet{Terasawa05}, \citet{Schwartz05}, and \citet{Tanaka07} after a few $e$-folding times. The latter reference also finds a similar break in the SGR 1900+14 giant flare. 

Note that the assumption of a constant reconnection rate for the duration of the hard spike ($\tau_{\rm spike}\sim0.1-1$ s), as applied in Section \ref{subsec:ec}, is suspect in the light of nonlinearity of the mode; the obtained estimates for $r$ and $\delta$ [Eq.~(\ref{eq:e conserv r}) and Eq.~(\ref{eq:e conserv delta})] are lower limits in this regard.

\subsubsection{Coalescence and impulsive bursty reconnection}\label{ssec:coalescence instability}

The least stable long-wavelength tearing modes ($\lambda^{\rm max}\sim L_y$) tend to saturate soon after the onset of the nonlinear phase. For efficient reconnection to occur the presence of a significantly strong external driver (e.g. sudden crustal motions at the footpoints of sheared arcades, the onset of an ideal instability, or the catastrophic ejection of a flux rope) may be required, which forces a current sheet to become unstable to shorter wavelength modes ($\lambda^{\rm max}\ll L_y$), such that a chainlike structure of magnetic islands is formed before the nonlinear phase sets in \citep{Uzdensky14}. This configuration is consequently unstable to the coalescence instability, whereby the magnetic islands approach each other through mutually attractive Lorentz forces since they essentially comprise parallel flowing current concentrations. Island coalescence is typically subdivided into two phases: (1) the ideal MHD phase, where the current loops approach one another, and (2) the resistive reconnection phase, where due to finite resistivity ($\eta\neq0$) and large field gradients between the approaching current loops, the loops merge to form one current loop with an increased cross section, i.e. larger magnetic island. Stability analysis was performed by \citet{Finn77} on a particular periodic island-chain configuration described by a Fadeev force-free equilibrium \citep{Fadeev65},
\begin{align}\label{eq:}
&\psi_0=\ln[\cosh(kx)+\epsilon\cos(ky)],\nonumber\\
&\mathbf{B}_0=B_0 \,\boldsymbol{\hat{z}}\times\boldsymbol{\nabla}\psi_0,\nonumber\\
&\nabla^2\psi_0=4\pi j_{z0}=(1-\epsilon^2)k^2\exp\left[-2\psi_0\right],\nonumber
\end{align}
where $\psi_0$ is the equilibrium magnetic flux function, $B_0$ the local (background) magnetic field, $j_{z0}$ the equilibrium current directed perpendicular to the reconnection plane, and $0<\epsilon<1$ the peakedness parameter of the current concentration in the magnetic islands. Subsequent numerical simulations have shown that, for a large range of $S_\delta$, the coalescence growth rate is much greater than the tearing growth rate \citep{Pritchett79} and that it depends critically on the value for $\epsilon$ \citep{Bhattacharjee83}; linear tearing corresponding to $\epsilon = 0$. 

The coalescence instability is characterised by two timescales associated with its distinct phases \citep{Kliem95}. During the ideal phase the current loops approach each other on a hydromagnetic timescale, whereby the length scale is given by the separation distance of paired current loops $\lambda_{\rm C}$, \begin{equation}\label{eq:}
\tau_{\rm C1}\sim\epsilon^{-1}\frac{\lambda_{\rm C}}{v_A}
\end{equation}
with $\delta\lesssim\lambda_{\rm C}\lesssim\lambda^{\rm max}$. In general $\tau_{\rm C1}\ll\tau_{\rm tm}^{\rm min}$, yet no magnetic energy is dissipated in the process. During the resistive phase, when the current loops merge, `anti-reconnection' occurs in-between the approaching islands. The reconnection rate is enhanced by the external driving forces of the converging current loops, such that in general $\tau_{\rm C2}<\tau_{\rm tm}^{\rm min}$. Moreover, for strongly peaked current concentrations ($\epsilon\to1$), we have $\tau_{\rm C1}\sim\tau_{\rm C2}\ll\tau_{\rm tm}^{\rm min}$. For $\lambda_{\rm C}^{\rm max}\simeq\lambda^{\rm max}\sim10^6$ cm, the coalescence timescale becomes comparable to the magnetospheric light crossing time, i.e. $\tau_{\rm C}\sim \tau_A^{\rm ext}\,\epsilon^{-1}$ ms.

Coalescence following tearing converts the bulk of the free magnetic energy in the current sheet, such that the island growth phase may act as mere prelude to the explosive energy release of merging current loops \citep{Leboeuf82}. Its rapid development and ability to convert a significant fraction of magnetic energy argue in favour of coalescence, rather than tearing, as an explanation for the impulsive phase of flares \citep{Tajima82,Tajima87,Sakai87,Kliem95, Schumacher97}. The observed giant flare emission may therefore be a proxy of the nonlinear, rather than the linear, tearing phase. 

Furthermore, for higher values of $\mathcal{S}_\delta$ and $\sigma_{\rm m}$ a nonlinear process known as `impulsive bursty reconnection' may occur, whereby a cycle of slow tearing, rapid coalescence, current sheet thinning, and further secondary tearing (i.e the plasmoid instability) at an increased rate repeats successively \citep{Leboeuf82,Priest85,Uzdensky10}. Consequently, energy is released during separate coalescence events in a fragmentary and quasi-periodic manner. This process is advanced to explain the periodic temporal fine structure of hard X-ray (HXR) emission and coherent drifting radio bursts associated with discrete (bidirectional) electron beams observed during the impulsive phase of solar flares \citep{Aschwanden95,Kliem00,Karlicky04}. Quasi-periodic oscillations (QPOs) of $\nu\sim10^2$ Hz, that might be associated with separate energy injections, have also been detected during the initial phases of the magnetar giant flares: See Subsection \ref{ssec:observed timescales}. These distinct energy surges may be interpreted as quasi-periodic peaks in coalescence rates\footnote{Quasi-periodic pulsations (QPPs) are ubiquitously observed in solar flares; among self-oscillatory reconnection, a multitude of alternative mechanisms have been proposed to explain these phenomena \citep{Nakariakov09}.}, resulting from impulsive bursty reconnection. Precise timing observations of hypothesised (drifting) radio burst from magnetars \citep{Lyutikov02} may greatly help to further probe the reconnection substructure (e.g. separate plasmoids), reconnection rate, and density of the acceleration region. Yet even though various magnetars show radio emission \citep[e.g.][]{Camilo06}, no such bursts of coherent radio emission coincident with the (recurrent) $\gamma$-ray bursts have been observed to date.

The total energy release through a multitude of coalescence events may be estimated accordingly \citep{Kruger89,Kliem95},
\begin{equation}\label{eq:}
U_{\rm C}^{\rm tot}\simeq\frac{(N_{\rm C}-1)^3}{N_{\rm C}}\frac{\lambda_{\rm C}^2L_zB_0^2}{24\pi^2}\ln\left(\frac{\lambda_{\rm C}}{\delta}\right),
\end{equation}
where $N_{\rm C}$ is the number of individual coalescence events. If we estimate the total number of coalescence events during the impulsive phase of a giant flares as follows,
\begin{equation}\label{eq:}
N_{\rm C}\simeq\nu\tau_{\rm spike}\frac{L_y}{\lambda_{\rm C}^{\rm max}}\sim10^2,
\end{equation}
we find for the total energy release through dynamic current sheet reconnection,
\begin{equation}\label{eq:}
U_{\rm C}^{\rm tot}\sim10^{45}\,N_{\rm C,2}^2(\lambda_{\rm C,6}^{\rm max})^2L_{z,6}B_{s,15}^2r_7^{-5}\ln\left(\frac{\lambda^{\rm max}_{\rm C,6}}{\delta_4}\right)~{\rm erg},
\end{equation}
where $L_{z,6}=L_z/10^6$ cm is the length of a current loop. This estimate is consistent with the observed energy output of the initial spike -- see Table \ref{tab:gf data}.

\section{Summary}

To better understand the extreme nature of the explosive onset of magnetar giant flares, we have discussed impulsive reconnection through the spontaneous development of the linear tearing instability in a globally sheared external field as a candidate trigger mechanism. Upon reexamination of previous works on the (relativistic) linear tearing mode, we found that the minimum growth time in magnetar magnetospheres is $\tau_{\rm tm}^{\rm min}\sim10^{-1}$ ms [Eq.~(\ref{eq:minimum growth time linear tearing mode})]. This estimate is consistent with the typical $e$-folding rise times ($\tau_e\sim0.1-1$ ms) of the giant flare light curves (see Table \ref{tab:gf data}). Our result differs significantly from the one found by \citet{Lyutikov03} ($\tau_{\rm tm}^{\rm L03}\sim10$ ms). Even though the rescaling of the current sheet thickness (by a factor of $10^{-2}$) has a larger effect on the final result, the difference is however essentially due to an error in that calculation.

Assuming the validity of the assumption that the exponential rise time of the giant flare is a proxy for the linear growth time of the tearing mode $\tau_{\rm tm}^{\rm min}=\tau_e$, we obtained order of magnitude estimates for the thickness of the current sheet and height of the base of the reconnection region, respectively $\delta\sim10^4$ cm and $r\sim10^7$ cm, through elementary pressure balance and energy conservation considerations. Additionally we found that the global length of the current sheet would have to be $L_y\sim(3-10)\,R_*$, which is reasonable for the giant flares, yet problematic for the smaller recurrent bursts, since such large unstable regions would have to develop on very short timescales $\Delta T\sim100$ s, where $\Delta T$ represents the typical waiting time of recurrent bursts \citep{Gogus99,Gogus00}. 

Finally we discussed the obtained constraints on the reconnection geometry and evaluated the soundness of the aforementioned assumption of equating an MHD growth time with an emission timescale. Regarding the latter, it is not apparent whether linear tearing dictates the rate of radiation release and if during the linear tearing phase magnetic field dissipation occurs efficiently enough to generate the observed emission. Considering the impulsive phase of solar flares there is no unequivocal connection between linear tearing and the observed high energy emission that is ultimately radiated by accelerated nonthermal particles. Moreover, substantial pair production in magnetar magnetospheres may obscure the emission resulting from magnetic field dissipation through reconnection, altogether. 

Furthermore, nonlinear effects become significant soon after the onset of linear tearing and in general reduce the reconnection rate considerably. Fast and efficient reconnection during the nonlinear impulsive bursty regime that may follow tearing, requires however the presence of a strong external driver e.g. rapid crustal motion or catastrophic loss of equilibrium of external magnetic field configurations. Accordingly, we propose that future research into magnetospheric trigger mechanisms for magnetar (giant) bursts investigate \emph{driven} reconnection scenarios, where the emission timescales may constrain the development of the external driver, the nonlinear reconnection phase, or the intense reconnection aftereffects.\\
\\
\textbf{Acknowledgments:} C.E. acknowledges support from NOVA (Nederlandse Onderzoeksschool voor Astronomie). A.W. acknowledges support from NWO Vidi Grant 639.042.916. The work of R.T. is partially supported by INAF through a PRIN grant. J.H. is supported by the Natural Sciences and
Engineering Research Council of Canada. We would like to thank Sam Lander, Lyndsay Fletcher, Maxim Lyutikov, Serguei Komissarov, and the participants of the `Integrated Plasma Modelling of Solar Flares' Lorentz Center workshop (May 2015) for helpful discussions. We also wish to acknowledge the useful and significant feedback from the anonymous referees.

\clearpage

\setlength{\topmargin}{2in}


\footnotesize

\begin{thebibliography}{54}
\expandafter\ifx\csname natexlab\endcsname\relax\def\natexlab#1{#1}\fi
\expandafter\ifx\csname url\endcsname\relax
  \def\url#1{{\tt #1}}\fi
\expandafter\ifx\csname urlprefix\endcsname\relax\def\urlprefix{URL }\fi

\bibitem[{{Aschwanden} et~al.(1995){Aschwanden}, {Benz}, {Dennis} et~al.}]{Aschwanden95}{Aschwanden} M.J., {Benz} A.O., {Dennis} B.R., et~al., 1995, \apj, 455, 347

\bibitem[{{Aschwanden}(2006)}]{Aschwanden06}{Aschwanden} M.J., 2006, Physics of the Solar Corona (Springer \& Praxis Publishing Ltd, Chichester, UK)

\bibitem[{{Barat} et~al.(1983){Barat}, {Hayles}, {Hurley} et~al.}]{Barat83}{Barat} C., {Hayles} R.I., {Hurley} K., et~al., 1983, \aap, 126, 400

\bibitem[{{Beloborodov} \& {Thompson}(2007){Beloborodov} \& {Thompson}}]{Beloborodov07}{Beloborodov} A.M., {Thompson} C., 2007, \apj, 657, 967

\bibitem[{{Beloborodov}(2009)}]{Beloborodov09}{Beloborodov} A.M., 2009, \apj, 703, 1044

\bibitem[{{Bhattacharjee} et~al.(1983){Bhattacharjee}, {Brunel}, \& {Tajima}}]{Bhattacharjee83}{Bhattacharjee} A., {Brunel} F., {Tajima} T., 1983, Phys. of Fluids, 26, 3332

\bibitem[{{Bibby} et~al.(2008){Bibby}, {Crowther}, {Furness}, {Clark}}]{Bibby08}{Bibby} J.L., {Crowther} P.A., {Furness} J.P., {Clark} J.S., 2008, \mnras, 386, L23 

\bibitem[{{Blaes} et~al.(1989){Bleas},{Blandford},{Goldreich}, \& {Madau}}]{Blaes89}{Blaes} O., {Blandford} R., {Goldreich} P., {Madau} P., 1989, \apj, 343, 839

\bibitem[{{Braithwaite} \& {Spruit}(2006)}]{Braithwaite06}{Braithwaite} J., {Spruit} H.C., 2006, \aap, 450, 1097

\bibitem[{{Braithwaite}(2009)}]{Braithwaite09}{Braithwaite} J., 2009, \mnras, 397, 763

\bibitem[{{Camilo} et~al.(2006){Camilo}, {Ransom}, {Halpern} et~al.}]{Camilo06}{Camilo} F., {Ransom} S.M., {Halpern} J.P., et~al., 2006, \nat, 442, 892

\bibitem[{{Cerutti} et~al.(2014){Cerutti},{Werner},{Uzdensky} \& {Begelman}}]{Cerutti14}{Cerutti} B., {Werner} G.R., {Uzdensky} D.A., {Begelman} M.C., 2014, \apj, 782, 104

\bibitem[{{Ciolfi} \& {Rezzolla}(2013){Ciolfi} \& {Rezzolla}}]{Ciolfi13}{Ciolfi} R., {Rezzolla} L., 2013, \mnras, 435, L43

\bibitem[{{Cline} et~al.(1980){Cline}, {Desai}, {Pizzichini} et~al.}]{Cline80}{Cline} T.L., {Desai} U.D., {Pizzichini} G., et~al., 1980, \apj, 237, L1

\bibitem[{{Davies} et~al.(2009){Davies}, {Figer}, {Kudritzki} et~al.}]{Davies09}{Davies} B., {Figer} D.F., {Kudritzki} R.-P., et~al., 2009, \apj, 707, 844

\bibitem[{{Duncan}(2004)}]{Duncan04}{Duncan} R.C., 2004, In: {H\"{o}flich} P., {Kumar} P., {Wheeler} J.C., 2004, (eds.) Cosmic Explosions in Three Dimensions, (Cambridge, UK: Cambridge Univ. Press)

\bibitem[{{Egedal} et~al.(2012){Egedal}, {Daughton}, \& {Le}}]{Egedal12}{Egedal} J., {Daughton} W., {Le} A., 2012, Nature Phys. Lett., 8, 321 

\bibitem[{{Fadeev} et~al.(1965){Fadeev}, {Kvabtskhava}, \& {Komarov}}]{Fadeev65}{Fadeev} V.M., {Kvabtskhava} I.F., {Komarov} N.N., 1965, Nucl. Fusion, 5, 202

\bibitem[{{Fenimore} et~al.(1996){Fenimore}, {Klebesadel}, {Laros}}]{Fenimore96}{Fenimore} E.E., {Klebesadel} R.W., {Laros} J.G., 1996, \apj, 460, 964

\bibitem[{{Feroci} et~al.(2001){Feroci}, {Hurley}, {Duncan}, {Thompson}}]{Feroci01}{Feroci} M., {Hurley} K., {Duncan} R.C., {Thompson} C., 2001, \apj, 549, 1021

\bibitem[{{Ferrario} \& {Wickramasinghe}(2006)}]{Ferrario06}{Ferrario} L., {Wickramasinghe} D., 2006, \mnras, 367, 1323

\bibitem[{{Finn} \& {Kaw}(1977){Finn} \& {Kaw}}]{Finn77}{Finn} J.M., {Kaw} P.K., 1977, 20, 72

\bibitem[{{Fletcher} \& {Hudson}(2008){Fletcher} \& {Hudson}}]{Fletcher08}{Fletcher} L., {Hudson} H.S., 2008, \apj, 675, 1645

\bibitem[{{Flowers} \& {Ruderman}(1977){Flowers} \& {Ruderman}}]{Flowers77}{Flowers} E., {Ruderman} M.A., 1977, \apj, 215, 302

\bibitem[{{Furth} et~al.(1963){Furth}, {Killeen}, \& {Rosenbluth} }]{Furth63}{Furth} H.P., {Killeen} J., {Rosenbluth} M.N., 1963, Phys. of Fluids, 6, 459

\bibitem[{{Gavriil} et~al.(2004){Gavriil}, {Kaspi}, \& {Woods}}]{Gavriil04}{Gavriil} F.P., {Kaspi} V.M., {Woods} P.M., 2004, \apj, 607, 959 

\bibitem[{Gedalin}(1996){Gedalin}]{Gedalin96}{Gedalin} M., 1996, \prl, 76, 3340

\bibitem[{Gill} \& {Heyl}(2010){Gill} \& {Heyl}]{Gill10}{Gill} R., {Heyl} J., 2010, \mnras, 407, 1926

\bibitem[{Goedbloed} et~al.(2009){Goedbloed}, {Keppens}, \& {Poedts}]{Goedbloed09}{Goedbloed} J.P., {Keppens} R., {Poedts} S., 2009, Advanced Magnetohydrodynamics (Cambridge: Cambridge Univ. Press)

\bibitem[{{G\"o\u{g}\"u\c{s}} et~al.(1999){Gogus}, {Woods}, {Kouveliotou} et~al.}]{Gogus99}{G\"o\u{g}\"u\c{s}} E., {Woods} P.M., {Kouveliotou} C., et~al., 1999, \apj, 526, L93

\bibitem[{{G\"o\u{g}\"u\c{s}} et~al.(2000){Gogus}, {Woods}, {Kouveliotou} et~al.}]{Gogus00}{G\"o\u{g}\"u\c{s}} E., {Woods} P.M., {Kouveliotou} C., et~al., 2000, \apj, 532, L121

\bibitem[{{G\"o\u{g}\"u\c{s}} et~al.(2001){Gogus}, {Kouveliotou}, {Woods} et~al.}]{Gogus01}{G\"o\u{g}\"u\c{s}} E., {Kouveliotou} C., {Woods} P.M., et~al., 2001, \apj, 558, 228

\bibitem[{Goldreich} \& {Reisenegger}(1992){Goldreich} \& {Reisenegger}]{Goldreich92}{Goldreich} P., {Reisenegger} A., 1992, \apj, 395, 250

\bibitem[{{Goldston} \& {Rutherford}(1995)}]{Goldston95}{Goldston} R.J., {Rutherford} P.H., 1995, Introduction to Plasma Physics. CRC Press.

\bibitem[{{Gourgouliatos} \& {Cumming}(2014){Gourgouliatos} \& {Cumming}}]{Gourgouliatos14}{Gourgouliatos} K.N., {Cumming} A., 2014, Phys. Rev. Let., 112, 171101

\bibitem[{{Harding} \& {Lai}(2006){Harding} \& {Lai}}]{Harding06}{Harding} A.K., {Lai} D., 2009, Rep. Prog. Phys., 69, 2631

\bibitem[{{Haschke} et~al.(2012){Haschke}, {Grebel}, \& {Duffau}}]{Haschke12}{Haschke} R., {Grebel} E.K., {Duffau} S., 2012, \apj, 144, 107

\bibitem[{{Henriksson} \& {Wasserman}(2013){Henriksson} \& {Wasserman}}]{Henriksson13}{Henriksson} K.T., {Wasserman} I., 2013, \mnras, 431, 2986

\bibitem[{{Heyl} \& {Kulkarni}(1998){Heyl} \& {Kulkarni}}]{Heyl98}{Heyl} J.S., {Kulkarni} S.R., 1998, \apj, 506, L61

\bibitem[{{Hoffman} \& {Heyl}(2012){Hoffman} \& {Heyl}}]{Hoffman12}{Hoffman} K., {Heyl} J., 2012, \mnras, 426, 2404

\bibitem[{Horowitz} \& {Kadau}(2009){Horowitz} \& {Kadau}]{Horowitz09}{Horowitz} C.J., {Kadau} K., 2009, \prl, 102, 191102

\bibitem[{{Hoshino} \& {Lyubarsky}(2012){Hoshino} \& {Lyubarsky}}]{Hoshino12}{Hoshino} M., {Lyubarsky} Y., 2012, Space Sci. Rev., 173, 521

\bibitem[{{Huang} \& {Yu}(2014a){Huang} \& {Yu}}]{Huang14a}{Huang} L., {Yu} C., 2014, \apj, 784, 168

\bibitem[{{Huang} \& {Yu}(2014b){Huang} \& {Yu}}]{Huang14b}{Huang} L., {Yu} C., 2014, \apj, 796, 3

\bibitem[{{Hurley} et~al.(1999){Hurley}, {Cline}, {Mazets} et~al.}]{Hurley99}{Hurley} K., {Cline} T., {Mazets} E., et~al., 1999, \nat, 397, 41

\bibitem[{{Hurley} et~al.(2005){Hurley}, {Boggs}, {Smith} et~al.}]{Hurley05}{Hurley} K., {Boggs} S.E., {Smith} D.M., et~al., 2005, \nat, 434, 1098

\bibitem[{Jones}(2003){Jones}]{Jones03}{Jones} P.B., 2003, \apj, 595, 342

\bibitem[{{Kagan} et~al.(2013){Kagan}, {Milosavljevic}, \& {Spitkovsky}}]{Kagan13}{Kagan} D., {Milosavljevic} M., {Spitkovsky} A., 2013, \apj, 774, 41

\bibitem[{{Karlicky}(2004)}]{Karlicky04}{Karlicky} M., 2004, \aap, 417, 325

\bibitem[{{Kliem}(1995)}]{Kliem95}{Kliem} B., 1995, Proc-1995-Benz, 93

\bibitem[{{Kliem} et~al.(2000){Kliem}, {Karlicky}, \& {Benz}}]{Kliem00}{Kliem} B., {Karlicky} M., {Benz} A.O., 2000, \aap, 360, 715

\bibitem[{{Komissarov}(2002){Komissarov}}]{Komissarov02}{Komissarov} S.S., 2002, \mnras, 336, 759 

\bibitem[{{Komissarov} et~al.(2007){Komissarov}, {Barkov}, \& {Lyutikov}}]{Komissarov07}{Komissarov} S.S., {Barkov} M., {Lyutikov} M., 2007, \mnras, 374, 415 (KBL07)

\bibitem[{{Kr\"uger} et~al.(1989){Kr\"uger}, {Kliem}, \& {Hildebrandt}}]{Kruger89}{Kr\"uger} A., {Kliem} B., {Hildebrandt} J., 1989, ESA-SP, 285, 169

\bibitem[{{Lander} \& {Jones}(2012){Lander} \& {Jones}}]{Lander12}{Lander} S.K., {Jones} D.I., 2012, \mnras, 424, 482

\bibitem[{{Lander}(2014)}]{Lander14}{Lander} S.K., 2014, \mnras, 437, 424

\bibitem[{{Lander} et~al.(2015){Lander}, {Andersson}, {Antonopoulou} et~al.}]{Lander15}{Lander} S.K., {Andersson} N., {Antonopoulou} D., et~al., 2015, \mnras, 449, 2047

\bibitem[{{Leboeuf} et~al.(1982){Leboeuf}, {Tajima}, \& {Dawson}}]{Leboeuf82}{Leboeuf} J.N., {Tajima} T., {Dawson} J.M., 1982, Phys. of Fluids, 25, 784 

\bibitem[{{Levin} \& {Lyutikov}(2012){Levin} \& {Lyutikov}}]{Levin12}{Levin} Y., {Lyutikov} M., 2012, \mnras, 427, 1574 

\bibitem[{{Link}(2014)}]{Link14}{Link} B., 2014, \mnras, 441, 2676

\bibitem[{{Liu} et~al.(2008){Liu}, {Petrosian}, {Dennis} et~al.}]{Liu08}{Liu} W., {Petrosian} V., {Dennis} B.R., {Jiang} Y.W., 2008, \apj, 676, 704

\bibitem[{{Liu} \& {Fletcher}(2009){Liu} \& {Fletcher}}]{Liu09}{Liu} S., {Fletcher} L., 2009, \apj, 701, L34

\bibitem[{{Low}(1973){Low}}]{Low73}{Low} B.C., 1973, \apj, 181, 209

\bibitem[{{Lyutikov}(2002)}]{Lyutikov02}{Lyutikov} M., 2002, \apj, 580, L65

\bibitem[{{Lyutikov}(2003)}]{Lyutikov03}{Lyutikov} M., 2003, \mnras, 346, 540 (L03)

\bibitem[{{Lyutikov}(2006)}]{Lyutikov06}{Lyutikov} M., 2006, \mnras, 367, 1594

\bibitem[{{Masada} et~al.(2010){Masada}, {Nagataki}, {Shibata} \& {Terasawa}}]{Masada10}{Masada} Y., {Nagataki} S., {Shibata} K., {Terasawa} T., 2010, \pasj, 62, 1093

\bibitem[{{Mazets} et~al.(1979){Mazets}, {Golenetskii}, {Il'Inskii} et~al.}]{Mazets79}{Mazets} E.P., {Golenetskii} S.V., {Il'Inskii} V.N., et~al., 1979, \nat, 282, 587

\bibitem[{{Mazets} \& {Golenetskii}(1981){Mazets} \& {Golenetskii}}]{Mazets81}{Mazets} E.P., {Golenetskii} S.V., 1981, Astrophys. \& Space Sci., 75, 47

\bibitem[{{Mazets} et~al.(1999){Mazets}, {Cline}, {Aptekar} et~al.}]{Mazets99}{Mazets} ., {Cline} ., {Aptekar} ., et~al., 1999, Astron. Lett., 25, 635

\bibitem[{{Mereghetti} et~al.(2005){Mereghetti}, {Tiengo}, {Esposito} et~al.}]{Mereghetti05}{Mereghetti} S., {Tiengo} A., {Esposito} P., et~al., 2005, \apj, 628, 938

\bibitem[{{Mereghetti} et~al.(2006){Mereghetti}, {Esposito}, {Tiengo} et~al.}]{Mereghetti06}{Mereghetti} S., {Esposito} P., {Tiengo} A., et~al., 2006, \apj, 653, 1423

\bibitem[{{Mereghetti}(2008)}]{Mereghetti08}{Mereghetti} S., 2008, \aapr, 15, 225

\bibitem[{{Miki\'c} \& {Linker}(1994){Miki\'c} \& {Linker}}]{Mikic94}{Miki\'c} Z., {Linker} J.A., 1994, \apj, 430, 898

\bibitem[{{Nakagawa} et~al.(2009){Nakagawa}, {Mihara}, {Yoshida} et~al.}]{Nakagawa09}{Nakagawa} Y.E., {Mihara} T., {Yoshida} A., et~al., 2009, Astron. Soc. of Japan, 61, 387

\bibitem[{{Nakariakov} \& {Melnikov}(2009){Nakariakov} \& {Melnikov}}]{Nakariakov09}{Nakariakov} V.M., {Melnikov} V.F., 2009, Space Sci. Rev., 149, 119

\bibitem[{{Narukage} \& {Shibata}(2006){Narukage} \& {Shibata}}]{Narukage06}{Narukage} N., {Shibata} K., 2006, \apj, 637, 1122

\bibitem[{{Olausen} \& {Kaspi}(2014){Olausen} \& {Kaspi}}]{Olausen14}{Olausen} S.A., {Kaspi} V.M., 2014, \apj Suppl. Series, 212, 6

\bibitem[{{Palmer} et~al.(2005){Palmer}, {Barthelmy}, {Gehrels} et~al.}]{Palmer05}{Palmer} D.M., {Barthelmy} S., {Gehrels} N., et~al., 2005, \nat, 434, 1107

\bibitem[{{Parfrey} et~al.(2013){Parfrey},{Beloborodov} \& {Hui}}]{Parfrey13}{Parfrey} K., {Beloborodov} A.M., {Hui} L., 2013, \apj, 774, 92

\bibitem[{{Petrosian} \& {Liu}(2004){Petrosian} \& {Liu}}]{Petrosian04}{Petrosian} V., {Liu} S.M., 2004, \apj, 610, 550

\bibitem[{{Priest}(1985)}]{Priest85}{Priest} E.R., 1985, IAU Symposia, 107, 233

\bibitem[{{Priest} \& {Forbes}(2000)}]{Priest00}{Priest} E., {Forbes} T., 2000, Magnetic Reconnection: MHD Theory and Applications (New York: Cambridge Univ. Press)

\bibitem[{{Pritchett} \& {Wu}(1979){Pritchett} \& {Wu}}]{Pritchett79}{Pritchett} P.L., {Wu} C.C., 1979, Phys. of Fluids, 22, 2140

\bibitem[{{Pucci} \& {Velli}(2013){Pucci} \& {Velli}}]{Pucci13}{Pucci} F., {Velli} M., 2013, \apjl, 780, L19

\bibitem[{{Rea} et~al.(2005){Rea}, {Tiengo}, {Mereghetti} et~al.}]{Rea05}{Rea} N., {Tiengo} A., {Mereghetti} S., et~al., 2005, \apj, 627, L133

\bibitem[{{Rutherford}(1973)}]{Rutherford73}{Rutherford} P.H., 1973, Phys. of Fluids, 16, 1903

\bibitem[{{Sakai} \& {Ohsawa}(1987){Sakai} \& {Ohsawa}}]{Sakai87}{Sakai} J.I., {Ohsawa} Y., 1987, Space Sci. Rev., 46, 113

\bibitem[{{Schumacher} \& {Kliem}(1997){Schumacher} \& {Kliem}}]{Schumacher97}{Schumacher} J., {Kliem} B., 1997, Phys. of Plasmas, 4, 3533

\bibitem[{{Schwartz} et~al.(2005){Schwartz}, {Zane}, {Wilson} et~al.}]{Schwartz05}{Schwartz} S.J., {Zane} S., {Wilson} R.J., et~al., 2005, \apj, 627, L129

\bibitem[{{Shibata} \& {Tanuma}(2001)}]{Shibata01}{Shibata} K., {Tanuma} S., 2001, Earth Planets Space, 53, 473

\bibitem[{{Shibata} \& {Magara}(2011){Shibata} \& {Magara}}]{Shibata11}{Shibata} K., {Magara} T., 2011, Living Rev. in Solar Phys., 8, 6

\bibitem[{{Sironi} \& {Spitkovsky}(2014){Sironi} \& {Spitkovsky}}]{Sironi14}{Sironi} L., {Spitkovsky} A., 2014, \apjl, 783, L21

\bibitem[{{Speiser}(1965)}]{Speiser65}{Speiser} T.W., 1965, J. of Geophys. Res., 70, 4219

\bibitem[{{Steinolfson} \& {van Hoven}(1984){Steinolfson} \& {van Hoven}}]{Steinolfson84}{Steinolfson} R.S., {van Hoven} G., 1984, Phys. of Fluids, 27, 1207

\bibitem[{{Tajima} et~al.(1982){Tajima},{Brunel}, \& {Sakai}}]{Tajima82}{Tajima} T., {Brunel} F., {Sakai} J., 1982, \apj, 258, L45 

\bibitem[{{Tajima} et~al.(1987){Tajima}, {Sakai}, {Nakajima} et~al.}]{Tajima87}{Tajima} T., {Sakai} J., {Nakajima} H., et~al., 1987, \apj, 321, 1031

\bibitem[{{Tanaka} et~al.(2007){Tanaka}, {Terasawa}, {Kawai} et~al.}]{Tanaka07}{Tanaka} Y.T., {Terasawa} T., {Kawai} N., et~al., 2007, \apj, 665, L55

\bibitem[{{Terasawa} et~al.(2005){Terasawa}, {Tanaka}, {Takei} et~al.}]{Terasawa05}{Terasawa} T., {Tanaka} Y.T., {Takei} Y., et~al., 2005, \nat, 434, 1110

\bibitem[{{Thompson} \& {Duncan}(1992){Thompson} \& {Duncan}}]{Thompson92}{Thompson} C., {Duncan} R.C., 1992, \apj, 392, L9

\bibitem[{{Thompson} \& {Duncan}(1995)}]{Thompson95}{Thompson} C., {Duncan} R.C., 1995, \mnras, 275, 255

\bibitem[{{Thompson} \& {Duncan}(1996){Thompson} \& {Duncan}}]{Thompson96}{Thompson} C., {Duncan} R.C., 1996, \apj, 473, 322

\bibitem[{{Thompson} \& {Duncan}(2001){Thompson} \& {Duncan}}]{Thompson01}{Thompson} C., {Duncan} R.C., 2001, \apj, 561, 980

\bibitem[{{Thompson} et~al.(2002){Thompson}, {Lyutikov}, \& {Kulkarni}}]{Thompson02}{Thompson} C., {Lyutikov} M., Kulkarni S.R., 2002, \apj, 574, 332

\bibitem[{{Tiengo} et~al.(2009){Tiengo}, {Esposito}, {Mereghetti} et~al.}]{Tiengo09}{Tiengo} A., {Esposito} P., {Mereghetti} S., et~al., 2009, \mnras, 399, L74

\bibitem[{{Turolla} et~al.(2015){Turolla}, {Zane}, \& {Watts}}]{Turolla15}{Turolla} R., {Zane} S., {Watts} A., 2015, Rep. on Progress in Phys., 73, 116901

\bibitem[{Uchida}(1997)]{Uchida97}{Uchida} T., 1997, Phys. Rev. E, 56, 2181

\bibitem[{{Uzdensky} et~al.(2010){Uzdensky}, {Loureiro}, {Schekochihin} et~al.}]{Uzdensky10}{Uzdensky} D.A., {Loureiro} N.F., {Schekochihin} A.A., et~al., 2010 , Phys. Rev. Lett., 105, 235002

\bibitem[{{Uzdensky}(2011)}]{Uzdensky11}{Uzdensky} D.A., 2011, \ssr, 160, 45

\bibitem[{{Uzdensky} \& {Loureiro}(2014){Uzdensky} \& {Loureiro}}]{Uzdensky14}{Uzdensky} D.A., {Loureiro} N.F., 2014, arXiv: 1411.4295

\bibitem[{{White}(1986)}]{White86}{White} R.B., 1986, Rev. Mod. Phys., 58, 183

\bibitem[{{Wolfson}(1995)}]{Wolfson95}{Wolfson} R., 1995, \apj, 443, 810

\bibitem[{{Woods} et~al.(1999){Woods}, {Kouveliotou}, {van Paradijs} et~al.}]{Woods99}{Woods} P.M., {Kouveliotou} C., {van Paradijs} J., et~al., 1999, \apj, 518, L103

\bibitem[{{Woods} et~al.(2001){Woods}, {Kouveliotou}, {G\"o\u{g}\"u\c{s}} et~al.}]{Woods01}{Woods} P.M., {Kouveliotou} C., {G\"o\u{g}\"u\c{s}} E., et~al., 2001, \apj, 552, 748

\bibitem[{{Yu}(2012)}]{Yu12}{Yu} C., 2012, \apj, 757, 67

\bibitem[{{Yu}(2013)}]{Yu13}{Yu} C., 2013, \apj, 771, L46

\bibitem[{{Zenitani} \& {Hoshino}(2001){Zenitani} \& {Hoshino}}]{Zenitani01}{Zenitani} S., {Hoshino} M., 2001, \apj, 562, L63

\bibitem[{{Zenitani} \& {Hoshino}(2007){Zenitani} \& {Hoshino}}]{Zenitani07}{Zenitani} S., {Hoshino} M., 2007, \apj, 670, 702

\bibitem[{{Zenitani} \& {Hoshino}(2008){Zenitani} \& {Hoshino}}]{Zenitani08}{Zenitani} S., {Hoshino} M., 2008, \apj, 677, 530

\end{thebibliography}
\end{document}